\documentclass[12pt]{iopart}

\usepackage{iopams} 
\usepackage{pdflscape}
\usepackage{longtable}
\usepackage{textgreek}

\usepackage[style=numeric,sorting=none,maxnames=5,minnames=3,doi=false,url=false,isbn=false,eprint=false]{biblatex}
\usepackage{footnote}
\usepackage{hyperref}
\usepackage{tablefootnote}

\usepackage{graphicx,caption,color}
\usepackage{wrapfig}

\newcommand{\naitl}{NaI(Tl)}
\newcommand{\nit}{N$_2$}

\newcommand{\mwe}{\,$\rm m.w.e.$\,}

\newcommand{\keVnr}{\mbox{keV$_{\rm{nr}}$}\,}
\newcommand{\eVnr}{\mbox{eV$_{\rm{nr}}$}\,}
\newcommand{\keVee}{\mbox{keV$_{\rm{ee}}$}\,}
\newcommand{\eVee}{\mbox{eV$_{\rm{ee}}$}\,}
\newcommand{\kevee}{keV$_{\rm{ee}}$\,}
\newcommand{\us}{$\mu$s}
\newcommand{\um}{$\mu$m}
\newcommand{\cms}{cm$^2$\,}
\newcommand{\cmt}{cm$^3$}
\newcommand{\tevc}{TeV/c$^2$}
\newcommand{\gevc}{GeV/c$^2$\,}
\newcommand{\mevc}{MeV/c$^2$\,}
\newcommand{\kevc}{keV/c$^2$\,}
\newcommand{\counts}{[kg~keV~day]$^{-1}$\,}
\newcommand{\kgyr}{[kg~yr]\,}
\newcommand{\kgd}{[kg~d]\,}
\newcommand{\tonyr}{[ton~yr]\,}
\newcommand{\tond}{[ton~d]\,}
\newcommand{\ppcge}{{\it p}PCGe}

\usepackage{color}

\begin{document}

\title[DM Annual Modulation Review]{Annual Modulation in Direct Dark Matter Searches}

\author[F. Froborg, A.R. Duffy]{Francis Froborg\dag, Alan R. Duffy\ddag}

\address{\dag Imperial College London, High Energy Physics, Blackett Laboratory, London SW7 2BZ, United Kingdom}
\address{\ddag Centre for Astrophysics and Supercomputing, Swinburne University of Technology, PO Box 218, Hawthorn, VIC 3122, Australia}
\ead{aduffy@swin.edu.au}
\vspace{10pt}
\begin{indented}
\item[]March 2020
\end{indented}

\begin{abstract}
The measurement of an annual modulation in the event rate of direct dark matter detection experiments is a powerful tool for dark matter discovery. Indeed, several experiments have already claimed such a discovery in the past decade. While most of them have later revoked their conclusions, and others have found potentially contradictory results, one still stands today. This paper explains the potential as well as the challenges of annual modulation measurements, and gives an overview on past, present and future direct detection experiments.
\end{abstract}

%
%
%
%
%

\section{Introduction}

There is ample evidence for the existence of dark matter in the Universe. The idea was initially inferred by the observation that rotation curves of spiral galaxies stay flat at large radii instead of declining as expected from the gravitational potential provided only from visible matter \cite{Babcock1939, Rubin1970}. 
The case for a dark component to large-scale structure in the Universe was dramatically strengthened, indeed sometimes  referred to as ``the smoking gun'' of dark matter, by the Bullet Cluster. In this event two colliding galaxy clusters revealed an offset between the centers of their baryonic mass (as traced by hot X-ray emitting gas) and their total mass as measured by gravitational lensing \cite{Markevitch2003}. 
This offset can then be used to infer how non-interacting, or collisionless, the dark matter is, albeit with some limitations~\cite{Robertson17}. 
On larger scales the fluctuations in the cosmic microwave background confirm  that only 4.8\% of the Universe is `ordinary' matter while the rest is 25.8\% dark matter, and 69\% dark energy \cite{Planck2015}. 
Not only do we know how much dark matter exists in the Universe, and approximately its collisionless nature, we can also trace out its filamentary nature on megaparsec scales using galaxies as visible tracers along the `Cosmic Web'~\cite{Percival01,Eisenstein05} and also directly through weak gravitational lensing~\cite{Massey07}. 
These large-scale structures are well explained by a dark matter term in both supercomputer simulations~\cite{Navarro97,Springel05,Schaye10} and analytic calculations of gravitational collapse theory~\cite{White91, Sheth01, Correa15b}.

However, it is still not clear what the nature of dark matter actually is. Initially, non-luminous massive compact halo objects (MACHOS) like brown / white dwarfs or even black holes as an explanation of the dark matter were proposed~\cite{Paczynski86}. 
However, it was found that these condensed baryonic objects can only be responsible for a fraction of the observed dark matter \cite{Tremaine1979, Pagel1992, Alcock00}. 
Just some of the current candidates include axions \cite{Peccei1977}, a postulated fourth species of `sterile' neutrinos~\cite{Boyarsky19} and the so-called Weakly Interacting Massive Particles (WIMPs). The latter is an umbrella term for particles that only interact through gravity and the weak force. 
They are theoretically motivated through supersymmetry \cite{Jungman1995} or other theories beyond the Standard Model (suggested reviews include\cite{Bergstrom2000, Bertone2004}) and are expected to have a mass from below \gevc to several \tevc.

There are three different possible processes to detect dark matter: astronomical signals from two dark matter particles self-annihilating (an indirect search), production at particle accelerators (seen as a `missing' amount of mass-energy from the resulting collision) or via scattering of a dark matter particle off a nuclei in a laboratory (direct searches). 
This review focuses on direct searches; for further information on accelerator searches or indirect searches, see for example \cite{Penning2017} or \cite{Gaskins2016} respectively. 
In the case of direct searches most experiments are counting experiments in the sense that they try to reduce or identify all potential background and then count potential events in their signal region. 
Another approach is the search for a modulation of the overall event rate in the region of interest. 
Such a modulation is expected due to the rotation of the Earth around the Sun. 
The rest of this review will focus on attempts to measure such a modulation.

In Section~\ref{sec:ER} we will explore the theoretical underpinning of the expected annual modulation signal, from the astrophysical considerations in Section~\ref{sec:ER_Astro} to particle physics effects in Section~\ref{sec:ER_Particle} and then the impact of experimental design on such a search in Section~\ref{sec:ER_Detector}. 
We then describe in detail the only claimed detection, albeit not without some controversy, for a dark matter annual modulation with the DAMA/LIBRA experiment in Section~\ref{sec:DAMA}. 
In Section~\ref{sec:ExpTests} we investigate range of current, and planned, detectors and categorize them into those that use the DAMA/LIBRA technology of sodium-iodide crystal to more directly investigate that claim (Section~\ref{sec:ExpTests_NaI}) and those that use other active detection techniques, such as noble gases or germanium (Section~\ref{sec:ExpTests_Other}).

\section{Event Rate and its Modulation}
\label{sec:ER}

Direct dark matter experiments search for a nuclear recoil caused by a WIMP $\chi$ elastically scattering off the target nuclei \cite{Goodman1984}.
The expected differential recoil rate per unit detector mass can be written as \cite{Freese2012}
\begin{equation}
\frac{dR}{dE_{\rm NR}} = \frac{1}{2m_\chi\mu^2}\,\sigma(q)\,\rho_\chi\,\eta (v_{\rm min}(E_{\rm NR}),t) 
\label{eqn:EventRate}
\end{equation}
where $E_{\rm NR}$ is the measured nuclear recoil energy, $\mu \equiv m_\chi\ M / (m_\chi + M)$ the reduced mass, $m_\chi$ the WIMP mass, $\sigma (q)$ the WIMP-nucleus cross section, $\rho_\chi$ the local dark matter mass density, and $\eta (v_{\rm min},t)$ the mean inverse speed. 
The advantage of writing the event rate in the form of Equation~(\ref{eqn:EventRate}) is that the particle physical ($\sigma(q)$) and astrophysical ($\rho_\chi\ \eta (v_{\rm min},t)$) components nicely separate. 
Each of these components contain some uncertainties and assumptions, which will be briefly discussed in the following, focussing on the impact on WIMP searches.

\subsection{Astrophysical Component}
\label{sec:ER_Astro}
For the local dark matter density $\rho_\chi$ and its velocity distribution $f(\bi{v})$, a smooth and well-mixed component with $\rho_\chi \approx 0.3$~GeV/\cmt\  and a spherical, isotropic, Maxwellian velocity distribution is typically assumed, the so-called Standard Halo Model (SHM). 
However, besides the isothermal sphere, other dark matter profiles like the NFW profile \cite{NFW} are suggested. 
Other potential deviations from the SHM include that the halo might instead be shaped in a more oblate or prolate way \cite{Katz1991, Dubinski1994, Debattista2008, Read2009}. 
Furthermore, the concentration of the dark matter halo, and hence overall local density, can depend on cosmology~\cite{Correa15a}. 
The gravitational effects of baryons can modify this concentration~\cite{Duffy10,Dutton16} and shape of the halo~\cite{Bryan13}. 
However, in practice these effects are minor deviations at the Solar Neighbourhood which lies inside the scale radius of the dark matter halo.

More significant possible deviations from the SHM include a so-called dark disk \cite{Lake1989, Read2008, Purcell2009, Ling2010, Pillepich2014}, or coherent flows in the dark matter from substructure like tidal streams (such from Sagittarius \cite{Purcell2012, Fiorentin2011}). 
Recent attempts to measure the local density include $\rho_\chi =0.542\pm0.042$~GeV/\cmt\ \cite{Bienayme2014} and $\rho_\chi =0.25\pm0.09$~GeV/\cmt\ \cite{Zhang2013}, which do not overlap in their uncertainties. 
New data such as from GAIA, as well as a reduction in assumptions\cite{Silverwood2015}, will be necessary to get a more accurate picture. 

The velocity distribution directly depends on the assumed density profile. 
The density distribution of the SHM is formally infinite, which directly relates to infinitely high velocities. 
This is clearly unphysical and typically taken care of by assuming a (still ad-hoc) smooth truncation of the velocity profile at the escape velocity of the galaxy \cite{Green2011}. 
This high velocity tail directly impacts WIMP searches, particularly so for those of low mass. 
Other uncertainties and assumptions on the density profile as discussed above directly translate into uncertainties on the local dark matter velocity relevant for WIMP searches. There are also concerns with the manner in which such numerical methods implement direct detection rate calculations~\cite{Green2010,Dblint2019}.  
For example, an anisotropic velocity distribution would result in a change of phase and shape of an annual modulation signal \cite{Green2003}, which will be discussed in more detail in section~\ref{sec:ER_AnnMod}. 

\subsection{Particle Physical Component}
\label{sec:ER_Particle}

For the coupling between the WIMP and the nucleus the simplest assumption is that the cross section is independent of momentum and velocity. In such scenarios, scalar or axial-vector couplings are assumed, which give rise to spin-independent (SI) and spin-dependent (SD) cross sections \cite{Jungman1995}, respectively. 
However, there are other well motivated interaction models. 
For example, dark matter with a magnetic dipole moment would result in a different nuclear response than SI and SD interactions \cite{Fitzpatrick2012, Gresham2014}. 
Furthermore, different mediators could be assumed, like the photon of electromagnetism, or by kinetic mixing of a massive gauge field with the photon \cite{Sigurdson2004, Banks2010, Chang2010, Barger2010}. 
It is also possible that the coupling of dark matter to protons and neutrons is different \cite{Feng2011}.

If the de Broglie wavelength of the momentum transfer becomes comparable to the size of the nucleus, the WIMP becomes sensitive to the spacial structure of the nucleus; and the cross section decreases with increasing momentum $q$ \cite{Lewin1996}. 
The nuclear form factor accounts for this effect, which differs depending on interaction model \cite{Helm1956, Bednyakov2006}. 
The Helm Form Factor $F$ also depends on the nuclear radius $r_n$ and recoil energy, with stronger significance for heavier target nuclei. 
Since this is a nuclear many-body problem, the form factors can only be approximated. 
This dependence of the interaction cross-section $\sigma$ can be represented by 
\begin{equation}
\sigma(qr_n) = \sigma_0F^2(qr_n), \label{eqn:sigma}
\end{equation}
where $\sigma_0$ is the zero-momentum transfer cross-section and $F$ can be modelled~\cite{Lewin1996} as a Bessel function $j_1$
\begin{equation}
F(qr_n) = 3\times \frac{j_1(qr_n)}{qr_n}e^{-(qs)^2/2}\label{eqn:helm}
\end{equation}
with $s$ a nuclear skin thickness and the effective nuclear radius $r_n = a_n A^{\frac{1}{3}} + b_n$ for a target of atomic number $A$ and $a_{n}, b_{n}$ in femtometers.

\subsection{Variations in the Event Rate}
\label{sec:ER_AnnMod}
\begin{figure}[t]
	\centering
	\includegraphics[width=0.7\textwidth]{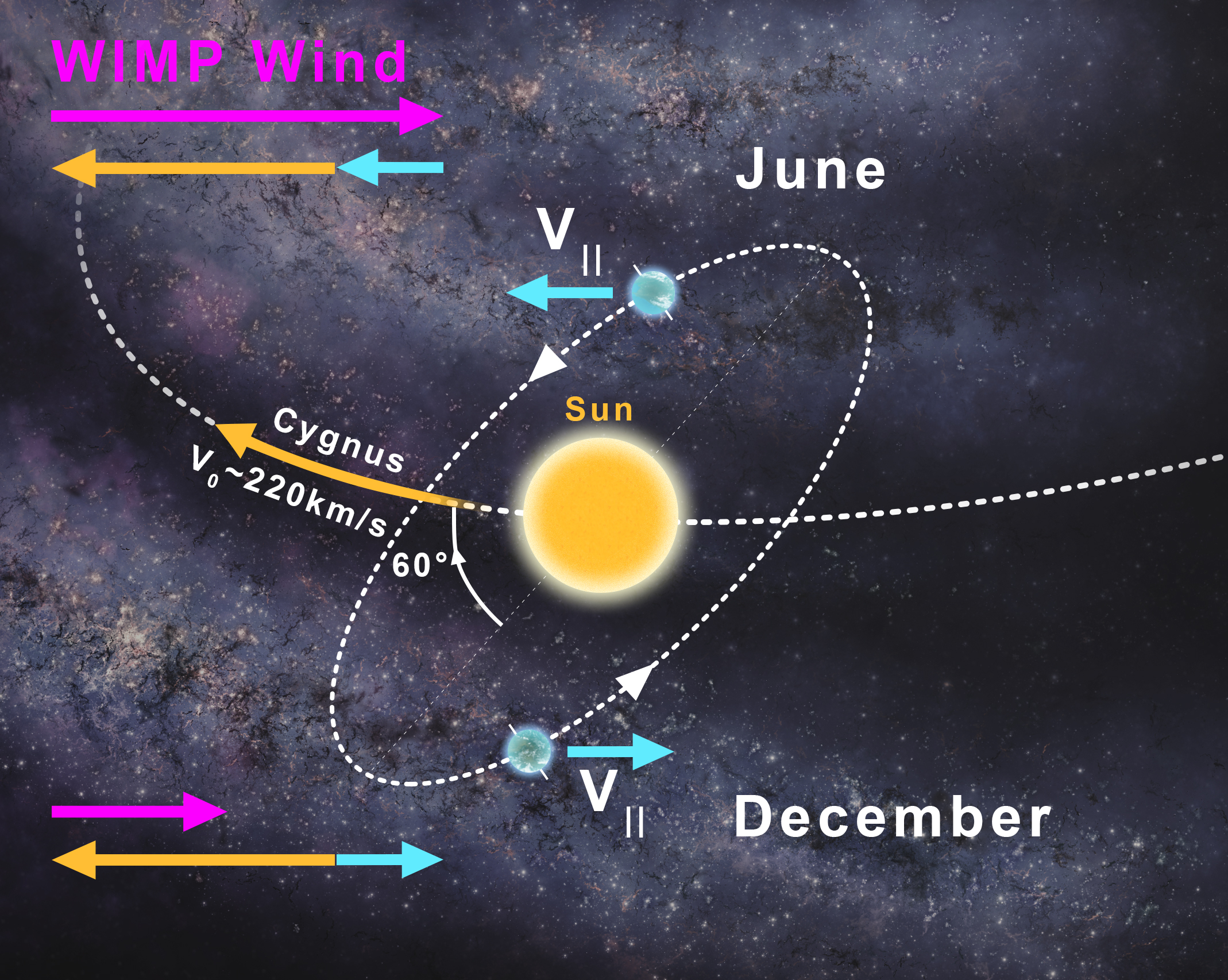}	\caption{Illustration of the rotation of the Earth around the Sun as well as the Sun around the galactic center, resulting in a WIMP wind with a predictable variation of intensity throughout the year. Reproduced with permission~\cite{SAP2019}.}
	\label{fig:wimpwind}
\end{figure}

Assuming a stationary halo, the rotation of the Sun around the Galactic center generates a constant dark matter ``wind" in the reference frame of the Sun. 
The strength of this wind will fluctuate on Earth due to its yearly rotation around the Sun. 
This effect, illustrated in Figure \ref{fig:wimpwind}, results in an annual modulation of the observed event rate. 
Because of the fixed period of one year, the differential scattering rate can be expanded in a Fourier series:
\begin{equation}
\frac{dR}{dE} (v_{\rm min} ,t) + A_0 + \sum_{n=1}^\infty A_n(v_{\rm min}) \cos\, n\omega(t-t_0) + \sum_{n=1}^\infty B_n(v_{\rm min})\sin\, n\omega (t-t_0)
\end{equation}
Assuming an isotropic and smooth halo component, this can be approximated by \cite{Freese2012}
\begin{equation}
\frac{dR}{dE} (E,t) \approx S_0(E) + S_m(E) \cos\,\omega(t-t_0)
\label{equ:modulation}
\end{equation}
with $|S_m|<< S_0$, where $S_0$ is the time-averaged rate, $S_m$ is the modulation amplitude, $\omega = 2\pi/{\rm year}$ and $t_0$ the phase of the modulation. 
Such a modulation would peak on June 1~\cite{Freese2012} and the amplitude is expected to be within a few percent of the constant WIMP contribution, depending on the halo model \cite{Drukier1986, Freese1988}. 
The modulating contribution is fairly weak due to an incline of $60^\circ$ between the Galactic plane and the ecliptic. 

The approximation in Equation~\ref{equ:modulation} might, however, be too simple. 
It is possible that the local dark matter is comprised of different dark matter components in which case neither the fixed phase nor sinusoidal shape of the modulation might be a good approximation \cite{Freese2012}. 
Furthermore, it might be possible that the dark matter particles are diverted on their path to Earth by the Sun's gravitational potential, effectively focussing the WIMP wind, which can result in a significant shift in the phase of the modulation \cite{Lee2013}.

Besides the annual modulation, there is also a component that varies on a daily basis due to the rotation of the Earth around its own axis. 
However, since the rotation velocity of the Earth (around 0.5~km/s at the equator) is significantly smaller than the orbital velocity (30~km/s), a signal from such a diurnal modulation will be significantly smaller than the annual modulation signal and much more difficult to detect.

\subsection{Experimental Approach}
\label{sec:ER_Detector}
The first published experiment attempting to detect an annual modulation signal used germanium at the Canfranc tunnel in the early nineties~\cite{DarkSide1994} and demonstrated the critical characteristics in such a search; sensitivity, stability and purity of detector, and shielding. An experiment needs to be sensitive enough to be able to measure the relatively small modulation contribution above the constant component from WIMPs and background. Background reduction and identification is typically reached by locating the experiment underground, usage of ultra-high purity materials, as well as active and passive shielding.
It is also crucial that the experimental conditions are stable and well-monitored to avoid that the signal is influenced by environmental effects like temperature, or pressure, which can have an effect on the detection efficiency. 
This typically includes an inner chamber that is sealed and filled with high-purity nitrogen gas, which is also necessary to avoid (varying) levels of radon close to the target. 
Radon is a noble gas and its concentration in outdoor air is typically around 10~Bq/m$^3$ \cite{WHO}; it decays through a chain of $\alpha$, $\beta$, and $\gamma$ emitters first into Pb-210 and then Pb-206. 

Since WIMPs by definition are electrically neutral, an interaction is only expected with the targets nucleus. 
Such a WIMP-nucleon interaction is expected in the low energy region. 
For example, a WIMP with a mass of about 100~\gevc\ will transfer roughly a few 10~keV of energy to the nucleus, depending on its velocity and the mass of the target material.
Thus, it is important for an experiment to have a high sensitivity to such low energies, a low threshold and a good energy resolution. Sensitivity in the low energy range gets even more important if you take into account that the recoiling nucleus transfers its energy either to electrons (observed e.g. via ionization or scintillation) or other nuclei (observed as phonons and heat). 
Since experiments are only sensitive to one or maybe two of these channels, they only observe a fraction of the total energy deposited, which is expressed in the quenching factor. 
Typical quenching factors are around 30\% \cite{Hitachi2005, SOMA2016, Cao2014, Xu2015b} but can be even lower as in the case of iodine scintillation experiments with a quenching factor of 9\% or potentially even lower \cite{Dama1996, Xu2015}. 
This can easily push the observed signal into the $<10$ \keVee region.

In the case of crystalline targets the channeling effect can also play a role: the scattered nucleus may recoil along the characteristic axis or plane of the crystal and thus travel fairly large distances without colliding with another nucleus. 
In this case nearly all the energy is transferred to electrons, resulting in a significant change of the quenching factor to $\approx 1$ \cite{Savage2009}. As pointed out originally by Lindhard, the channeling effect would not be present in a perfect crystal and in the absence of energy-loss processes \cite{Lindhard1965}. 
Thus, vibrations play a crucial role, resulting in a strong temperature dependency of the channeling effect.
Furthermore, channeling can in principle result in a modulation signal since the rotation of the Earth changes the orientation of the crystal structure towards the WIMP wind \cite{Bozorgnia2010}. 

Migdal pointed out that the rapid change in velocity of the recoiling nucleus can cause bound electrons to become excited or ionized \cite{Migdal1941}. 
Although the energy of these electrons can be transferred into observable modes there is a timing difference between this contribution and that of the recoiling nucleus transferring its energy to other nuclei (the standard signal). 
It is therefore possible that this energy is not properly taken into account and that the energy of the recoiling nucleus appears smaller. 
The Migdal effect is known to be smaller for targets with larger atomic number $Z$ due to the quadratic dependency of the Coulomb force, which results in a $1/Z^2$ proportionality of the Migdal effect \cite{Tomozawa2008}.

The advantage of modulation experiments above counting experiments is that they do not necessarily have to be able to identify all background events, as long as they have enough sensitivity to measure the expected WIMP modulation amplitude and can show that no potential background or other effect could produce a similar variation in the event rate.

\section{The DAMA Experiments and Their Result}
\label{sec:DAMA}

The first experiment that measured a signal compatible with a WIMP interaction is DAMA, which has measured a modulation in the event rate for over twenty years now. 
However, the interpretation as a WIMP signal is in tension with the null results of counting experiments such as LUX, XENON, or SuperCDMS \cite{Lux2017, Xenon20171t, Supercdms2014}.
DAMA does however effectively rule out a range of masses under the assumption of spin independent interactions~\cite{Baum2019}.

\subsection{The DAMA Experiments}

The DAMA experiment (initially called DAMA/NaI and later \cite{Dama1999} and DAMA/LIBRA \cite{Dama2008}) uses ultra-high purity \naitl\ crystals and measure the scintillation signal produced by electron and nuclear recoils. 
It is located underground at LNGS in Italy, which provides 3200\mwe of shielding to the top \cite{Bettini2007}. DAMA/NaI started with $\sim100$~kg of \naitl\ crystals, which were then upgraded to $\sim250$~kg for DAMA/LIBRA. 
In both cases the crystals are coupled to two photomultiplier tubes (PMTs) on each side of the crystal through a 10~cm synthetic quartz light guide. The setup is placed inside a copper box filled with high-purity \nit\ (radon box) and is passively shielded with layers of high-purity copper, lead, polyethylene, and concrete.

\subsection{The Modulation Signal}
\begin{figure}[t]
	\includegraphics[width=\textwidth]{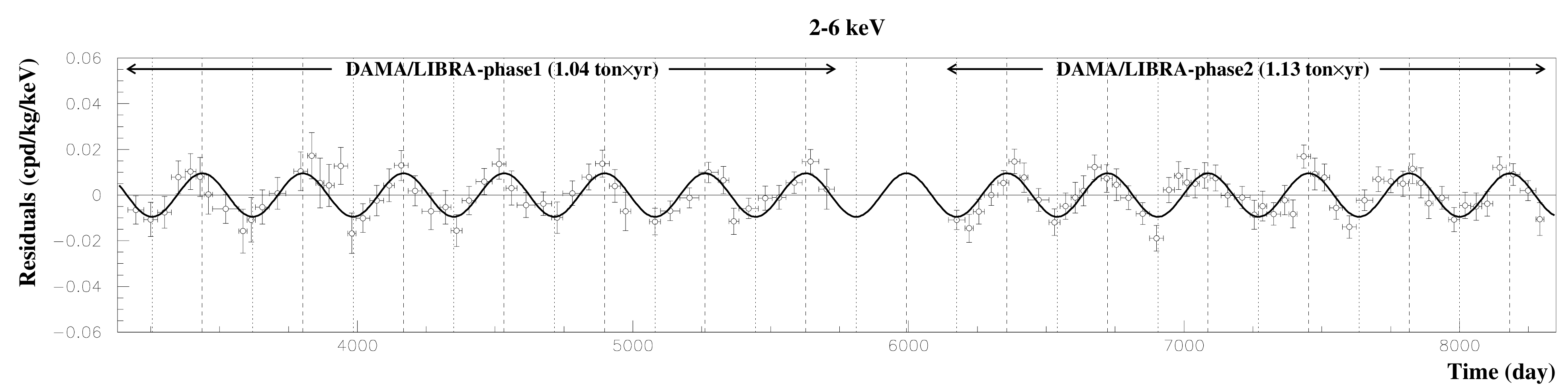}
	\caption{Variations in the event rate of the two DAMA/LIBRA phases from~\cite{Dama2018}. In both cases, only single interactions in the detectors are taken into account. The solid black curve is a sinusoidal fit assuming an annual modulation with parameters consistent with a WIMP interaction. Further details in the text.}
	\label{fig:dama-mod-all}
\end{figure}

Physics data was taken with DAMA/NaI from 1995 -- 2002 and with DAMA/LIBRA-phase I from 2003 -- 2010~\cite{Dama2013}, continuing with DAMA/LIBRA-phase II from 2011 onward~\cite{Dama2018}. 
A model independent analysis~\cite{Dama2018} of twenty years of data confirms a modulation of an event rate in the $(2-6)$~\keVee\ region over 14 annual cycles from DAMA/LIBRA phase I and II, as shown in Figure~\ref{fig:dama-mod-all}. 
A corollary analysis of this data~\cite{Dama2019} reduces the software energy threshold from $2$ to $1$~\keVee for the first six annual cycles and confirms the annual modulation at a $12.9 \sigma$ detection level.

The modulation has all the right features: the phase is $145\pm5$ days (2 June), the period is $T = 0.999\pm0.001$~yr, and it is only visible in the low energy region and only in single-hit events (meaning there was only one interaction in the entire detector at the time). 
Comparing the measured amplitude of the modulation of $S_m = (0.0103\pm0.0008)$~cpd/kg/\kevee\ to the total event rate in that energy region as published in \cite{Dama2008b}, the modulation is an $O(1\%)$ effect\footnote{Beware that the total event rate includes the constant WIMP contribution as well as the background in this energy region.}. 
With a total exposure of 2.46~ton$\times$yr, they measure the modulation with a 12.9$\sigma$~C.L. 
In the rest of this review this combined dataset will be referred to as DAMA. 

Due to the different masses of the two main target components WIMPs with two different masses could reasonably produce such a modulation: $m_\chi\sim10$~\gevc\ from Na recoils and $m_\chi\sim80$~\gevc\ from I recoils \cite{Savage2009, Dama2008b}. 
As pointed out earlier, the conversion from measured recoil energy into WIMP mass and WIMP-nucleon cross section involves a couple of assumptions and uncertainties. Notably in case of NaI, this includes uncertainties in the quenching factors, which have been measured by different groups \cite{Collar2013, Xu2015, Stiegler2017}. 
DAMA uses constant quenching factors of 0.3 for Na and 0.09 for I \cite{Dama1996} while the recent measurements indicate an energy dependency with smaller quenching factors in the energy region of interest. 
The fractions of events that undergo the channeling or Migdal effect (see Section~\ref{sec:ER_Detector}) can only be approximated (see e.g. \cite{Bozorgnia2010}). References \cite{Dama2007ch} and \cite{Dama2007mig} describe how DAMA takes both effects into account in their analysis.
Others have observed no such channeling effects in their measurements at all \cite{Collar2013, Xu2015}. 
However, the effect could potentially be somewhat dependent on the exact crystal structure and thus varies between different samples. 

The most significant enhancement to DAMA in Phase II has been the reduction in the energy threshold floor of detection, from 2 to 1 \keVee  (electron equivalent). 
This limit now rules out spin-independent dark matter models at 8 GeV by $5.2 \sigma$, and less strongly at higher masses, with 54 GeV disfavoured at the $2.6 \sigma$ level~\cite{Baum2019}. 
The additional freedom through spin dependent interactions does however permit good fits to low / high mass candidates of 10 GeV and 45 GeV respectively.

\section{Worldwide Detection Experiments}
\label{sec:ExpTests}

\begin{figure}[ht]
	\centering
	\includegraphics[width=\textwidth]{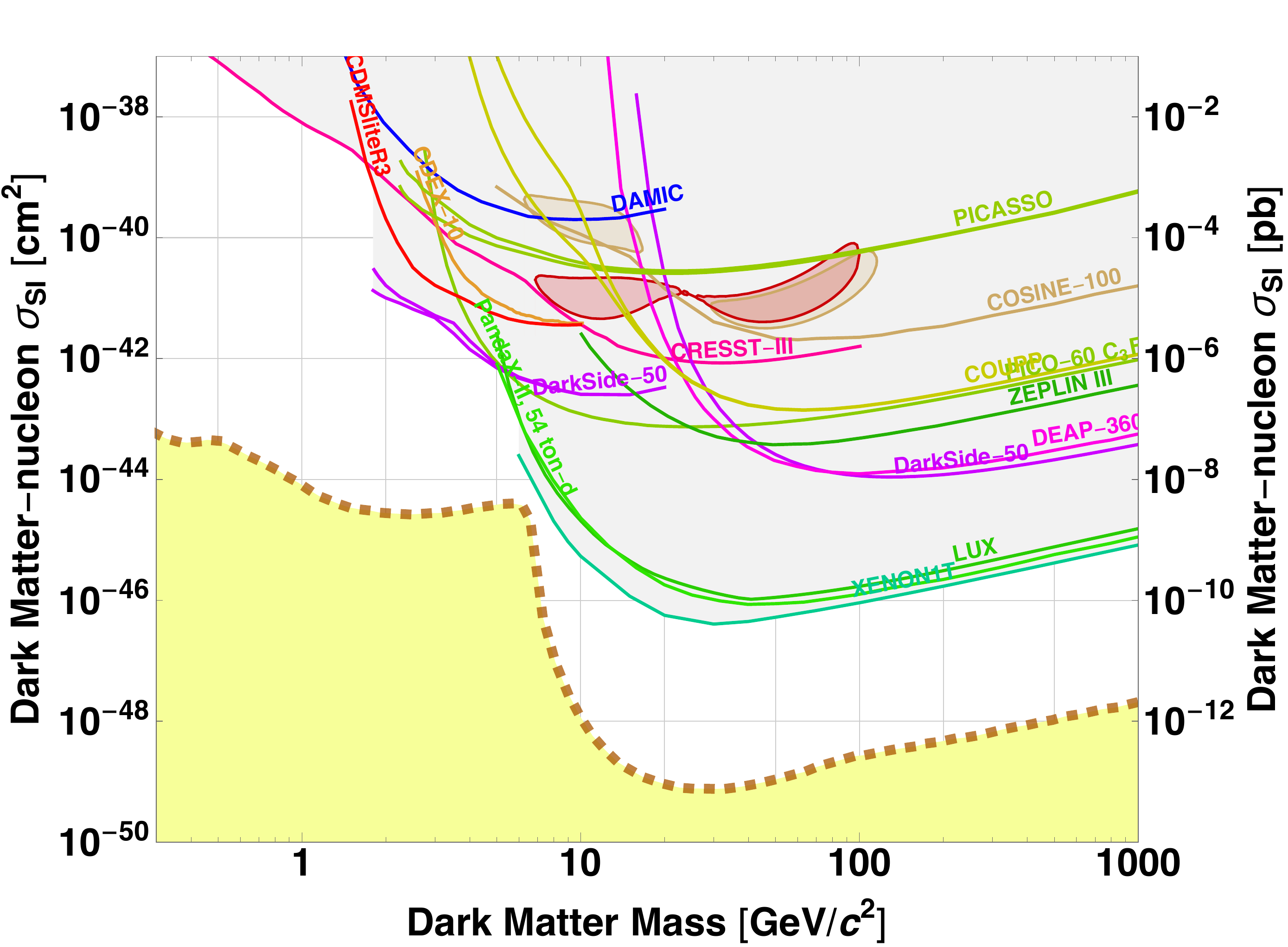}
	\caption{The current spin-independent interaction limits as a function of WIMP mass, as generated using SuperCDMS Dark Matter Limit Plotter \url{https://supercdms.slac.stanford.edu/dark-matter-limit-plotter}. The reddish (mustard) circular region is the claimed detection by DAMA with (without) ion channel effects considered, everything else is an exclusion limit for experiments outlined in the main text. The lower yellow limit is the Solar neutrino floor with a Xenon target. We have not included all experiments described in the main text, or even progressive upgrades of certain experiments that {\it are} shown in this image, to reduce the  overlapping curves in this plot. We have also chosen not to present the predicted limits for next generation detectors that are described in the text for the same reason.}
	\label{fig:dmlimits}
\end{figure}

The search for the annual modulation signal of dark matter has now spanned the globe, with most advanced nations hosting (or contributing) to at least one experiment.
Several of experiments have published confirming, as well as contradicting results, of DAMA. 
Some of these claims still stand today, some later revoked. We have summarised the constraints from these experiments in Fig~\ref{fig:dmlimits}. 

This section gives an overview on those experiments that at some point published a positive result that could be interpreted as elastic scattering of a WIMP on a target nucleus. 
The descriptions of the following experiments as well as their data analysis and results are kept brief and details can be found in the corresponding publications. 
We have structured these experiments into those adopt a technology design similar to DAMA/LIBRA, i.e. sodium-iodide crystals, for more direct comparison in Section~\ref{sec:ExpTests_NaI}. 
Then in Section~\ref{sec:ExpTests_Other} we explore those experiments using other active detector technologies, such as noble gases.
All experiments are summarized with their most relevant detector properties, sensitivities and timelines in Table~\ref{tab:detectors}.

\begin{landscape}
\begin{center}
    
\setlength\LTleft{0pt}
\setlength\LTright{0pt}

\begin{longtable}{p{37mm} p{20mm} p{20mm} l p{25mm} p{21mm} p{50mm} p{7mm}}
\caption{Overview on the different Direct Dark Matter experiments and their parameters sorted by target mass.} \\

\hline
Name        & Type      & Detector   & Exposure & Background    & Threshold   & Status  & Ref. \\ 
        &       & Mass [kg]    & \kgd  &  \counts   & \keVee    &   &  \\ \hline\hline
\endfirsthead

\hline
Name        & Type      & Detector   & Exposure & Background    & Threshold   & Status  & Ref. \\ 
\hline\hline
\endhead

\hline \multicolumn{8}{|r|}{{Continued on next page}} \\ \hline
\endfoot

\hline \multicolumn{8}{|r|}{Table End}\\ \hline
\endlastfoot

DAMA/LIBRA-I   & NaI(Tl)    & 232.8                   & 1.04 \tonyr     & 0.36                      & 2.0          & 7 year run complete                             & \cite{Dama2013}       \\ \hline
DAMA/LIBRA-II   & NaI(Tl)   & 242.5                   & 1.13 \tonyr  & 0.36                      & 1.0          & 6 year run complete                             & \cite{Dama2018}       \\ \hline
DM-ICE      & NaI(Tl)   & 17                    & 60.8 \kgyr    & 7.9                       & 6.5                  & 2011 - 2013                       & \cite{dmice2017}          \\ \hline
COSINE-100      & NaI(Tl)   & 106                   & 97.7 \kgyr             & 2.7                       & 2.0                  & Finished 2 yr run, upgrading to COSINE-200   & \cite{Cosine100b}          \\ \hline
KIMS       & CsI   & 103.4 (final)           & 67.2 \kgyr             & 0.11                   & 3 \footnote{equivalent of 2\keVee in Na}                 & Now COSINE              & \cite{Kims2012}          \\ \hline
KIMS - NaI  & NaI(Tl)   & 17.4          & N/A              &  3                      & $<$2.0                 & Now COSINE              & \cite{Kims2014,Kims2016}          \\ \hline
ANAIS       & NaI(Tl)   & 112.5                  & 220.69 \kgyr             & 3.6                       & 1.0                  & Five year run ongoing              & \cite{Amare2019}          \\ \hline
SABRE-PoP   & NaI(Tl)   & 5.2                   & N/A     & 0.36                      & 2.0         & Operating                             & \cite{SABRE_SABREPOP}       \\ \hline
SABRE       & NaI(Tl)   & 50                    & 150 \kgyr    & 0.1                       & $<$1.0                  & Est. Q1 2021                          & \cite{Shields2015}          \\ \hline

CoGeNT      & Ge       & 0.443 (0.33 fid.) & 373           & 1.82                      & 0.5                   & Upgrading to C-4                      & \cite{Cogent2014} \\ \hline

CDMS-II      & Ge      & 4.6 & 612      & 0.9 total                & 10                 & Upgraded to SuperCDMS                    & \cite{Cdms2010} \\ \hline

CDMS-II - Low E      & Ge      & 1.912 & 241      & $\sim 1.0$ (0.1)   & 2 (5)      & Upgraded to SuperCDMS                    & \cite{Cdms2011} \\ \hline

CDMS-II - Si   & Si      & 0.848 & 140.2 \footnote{equivalent to 23.4 \kgd over 7 -- 100 \keVee range for WIMP mass 10 \gevc after applying selection criteria } & $ \sim 0.5$ (total)             & 7                 & None                    & \cite{Cdms2013} \\ \hline

CDMSlite      & Ge      & 0.6 & 70.1       & 16.33 (1.09) & 0.056 (0.2)  & \parbox{45mm}{Upgraded to SuperCDMS-SNOLAB}                    & \cite{CDMSliteII} \\ \hline

SuperCDMS     & Ge      & 9 & 1690        & $\sim 1.0$ (total)               & 8                & \parbox{45mm}{Upgraded to SuperCDMS-SNOLAB}                    & \cite{Supercdms2018} \\ \hline

SuperCDMS-SNOLAB     & Ge      & 9 & 1690         & $\sim 1.0$ (total)               & 8                & Online 2020                & \cite{SupercdmsSNOLAB} \\ \hline

CDEX-1A     & Ge       & 0.994 (0.919 fid.) & 53.9         & 4.09                   & 0.475                   & Upgraded to CDEX-10                 & \cite{Cdex2014} \\ \hline

CDEX-1B     & Ge - NaI(Tl)       & 1.0 (0.939 fid.)  &  737.1           & $\sim 7\,(2.5)$                 & 0.16 (2.5)                  & Upgraded to CDEX-10                 & \cite{Cdex2017} \\ \hline

CDEX-10      & Ge        & 10 \footnote{Although 10 kg of detector mass was created, only a fiducial mass of 0.939 kg was ever successfully used in a live run} & 102.8            & 2.47                      & 0.16                 & Upgrading to CDEX-1T                  & \cite{Cdex10} \\ \hline

EDELWEISS-II      & Ge        & 4 (1.6 fid.) & 384           & $6 \times 10^{-4}$                     & 2.5 \keVnr                 & Upgraded to EDELWEISS-II                  & \cite{EdelweissII} \\ \hline

EDELWEISS-III      & Ge        & 20.9 (15.6 fid.) & 582            & $6 \times 10^{-4}$                     & 2.5 \keVnr                 & Upgrading to EURECA                  & \cite{EdelweissIIIb} \\ \hline

CRESST-II P1   & CaWO$_{4}$   & 10         & 730     & 42\footnote{Summation of all background events in M1 global maximum detection analysis gives 42 events, in the energy region of interest 10 - 40 \keVee}     & 10 - 19\footnote{Module specific cuts are imposed to limit background to 1 $\gamma$ per module per run}   & Completed 2011, upgraded to Phase 2  & \cite{Cresst2012}          \\ \hline

CRESST-II P2       & CaWO$_{4}$   & 0.249                & 29.35            & 3.51                     & 0.6                 & Completed 2014, upgraded to Phase 3           & \cite{Cresst2015}          \\ \hline

CRESST-III      & CaWO$_{4}$   & 0.024             & 3.64            & 1 per \kgd                     & 0.03                 & Phase III-1 completed 02/2018, Phase III-2 underway           & \cite{Cresst2019}          \\ \hline

DAMIC   & Si CCD  & $3 \times 0.0029$             & 0.6             & 30                     & 0.06                 & Upgrading to DAMIC100           & \cite{DAMIC}          \\ \hline

Zeplin-III  & Xe (dual phase) & 12 (6.5 fid.)            & 1344     & $<3 \times 10^{-4}$          & 7.4           & Upgrading to LUX-ZEPLIN         & \cite{ZeplinIIIb} \\ \hline 

LUX   & Xe (dual phase) & 370 (250 fid.)            & $3.35 \times 10^{4}$             & $1.7 \times 10^{-3}$          & 3 (0.7 possible)               & Upgrading to LUX-ZEPLIN            & \cite{Lux2017} \\ \hline

XENON100  & Xe (dual phase) & 370 (250 fid.)            & $4.77 \times 10^{4}$             & $2.6 \times 10^{-3}$          & 2              & Upgradeing to XENON1T  & \cite{Xenon100} \\ \hline

XENON1T  & Xe (dual phase) & 3200 (2000 fid.)            & 22 \tond             & $< 10^{-3}$          & 0.4              & Upgrading to XENONnT & \cite{Xenon2019a} \\ \hline

PandaXII  & Xe (dual phase) & 580 (330 -- 360 fid.) & 54 \tond             & $0.8 \times 10^{-3}$          & 1              & Upgrading to PandaX-4T & \cite{PandaXII} \\ \hline

XMASS  & Liquid Xe & 835 (832 fid.) & 1.8 \tonyr        & $0.75$          & 1              & Upgrading to XMASS-1.5 & \cite{Xmass2018} \\ \hline

XMASS (Low B) & Liquid Xe & 835 (97 fid.) & 68.5 \tond     & $4.2 \times 10^{-3}$          & 2              & Upgrading to XMASS-1.5 & \cite{Xmass2019} \\ \hline

DarkSide-50 & Ar (dual phase) & 46.4     & 2616             & Nil (after cuts)         & 13 \keVnr           & 3-yr run started & \cite{DarkSide16} \\ \hline

DarkSide-50 (Low Mass)  & Ar (dual phase) & 46.4     & 6786             & 1.5 (in entire run)         & 0.1              & Upgrading to DarkSide-20k & \cite{DarkSide18a,DarkSide18b} \\ \hline

DEAP-3600  & Liquid Ar & 3279     & 758 \tond             & Nil (after cuts)         & 15.6              & Upgrading to DarkSide-20k & \cite{Deap2019} \\ \hline

PICASSO  & C$_{4}$F$_{10}$ & 3      & 213.4            & Nil       & 1 \keVnr              & Upgraded to PICO-60  & \cite{PICASSO} \\ \hline

COUPP  & CF$_{3}$I & 4      & 437.4          & Nil       & 1 \keVnr              & Upgraded to PICO-60  & \cite{COUPP} \\ \hline

PICO-60  & C$_{3}$F$_{8}$ & 52      & 1404      & Nil       & 2.45          & Upgrading to PICO-500  & \cite{Amole19} \\ \hline

\label{tab:detectors}

\end{longtable}
\end{center}

\end{landscape}

\subsection{Sodium Iodide Crystal Detectors}
\label{sec:ExpTests_NaI}
This section describes sodium iodide experiments that try to directly test the results from DAMA. As the sole experimental claim of an annual modulation the focus of future experiments will be to test this result. 
Ideally this test is to be undertaken with the same detector type as spin-dependent interaction cross-sections for dark matter can accommodate a DAMA detection with null results from experiments using different elements, e.g. xenon, as well as channelling effects from the crystal lattice itself. 

Beyond providing the most direct comparison with the DAMA experiment a crystal made from NaI offers an incredibly low background material as impurities can be reduced by orders of magnitude, in theory, during the fabrication process. 
Furthermore, the target nucleon mass similar to the 10 - 100 \gevc WIMPs, best motivated from astrophysics considerations, which ensures a maximal kinematic coupling and hence energy transfer from colliding WIMP to recoiling nucleon.

To test the claims from the long-running DAMA experiment in a timely manner requires planned NaI-based detectors to combine both large detector mass as well as dramatically lower backgrounds. 
The latter is particularly important as nuclear recoils for NaI occur at an energy range dominated by environmental radioactive, seasonal and cosmic background~\cite{Urquijo:2016}. 
Utilising an underground lab with several kilometres of water equivalent depth overburden can suppress cosmic, i.e. muon, event rates by over 3 orders of magnitude as required for dark matter studies. 
The seasonal effects can only be tested by a dual hemisphere experiment, as proposed in SABRE discussed below, meaning shielding against local radioactive sources (and resultant gammas and spallation neutrons) are a key consideration. 

Even after placing the detector in an optimally shielded underground laboratory, well-ventilated to remove the build up of radon gas released by surrounding rock, NaI experiments are typically limited by the radioactive impurities in the crystal itself. 
Producing such ultra-clean crystals is immensely challenging, requiring bespoke ultra-pure initial powder and then superb control against possible contamination at each stage of growth. 
This includes ensuring the crucible themselves don't lead to accidental contamination. 
As impurities concentrate in the growing tip of the crystal this last part can be cut away, and the remaining crystal melted and regrown. A repetition of this process can lead to the required ppt contamination levels of $^{238}$U and $^{232}$Th and ppb levels of potassium~\cite{Xu2015}.

However, even at this level of contamination the NaI experiments are systematically limited by the background signal from remaining $^{40}$K impurities.
The decay chain for this isotope can produce Argon through electron capture which then emits a 1.46 ${\rm MeV}$ $\gamma$ and an Auger electron at a characteristic 3 \keVee \, energy.
This electron can be mistaken for a recoil event by dark matter and is the major source of background for DAMA. 
The detection of the $\gamma$ in a scintillation fluid surrounding the NaI can be used to screen against these `dark matter' events in coincidence. 
This so-called active veto system was first proposed in this context by the SABRE collaboration in 2013~\cite{Shields2015}.

\subsection{DM-ICE}
Uniquely among dark matter direct detection experiments, DM-ICE was deployed to the Antarctic making it also the first Southern Hemisphere detector. The 17kg of NaI(Tl) crystal was installed 2457m into the ice, co-located with IceCube, in 2010~\cite{dmice2014}. 
Although a relatively small mass, it was the first detector that utilised the same detector crystal type to DAMA providing the first opportunity to directly investigate their claimed detection. 
This crystal was planned to be the first of several, proving the feasibility to stably run a NaI(Tl) crystal in this environment. 
Unfortunately, no other crystals were ready to be deployed at the latest IceCube upgrade and as a consequent no other detectors could (yet) be installed.

The first two years of commissioning data, July 2011 -- June 2013, were sufficient to constrain the background in the 6.5 -- 8.0 \keVee\ energy range of $7.9\pm 0.4$ counts \counts which is consistent with the detector background itself.

\subsection{KIMS}

The Korea Invisible Mass Search (KIMS) experiment began in 2003 with one CsI(Tl) 6-kg crystal placed in the Yangyang Underground Laboratory in Korea.
Although formally a different crystal setup to the NaI of DAMA/LIBRA it is included in this Section as it can provide a direct test to the DAMA result. 
The setup uses passive and active shielding (liquid-scintillator-loaded mineral oil) and was gradually upgraded to first 4 and finally 12 crystals with a total mass of 103.4~kg. 

A significant run~\cite{Kims2012} using this setup achieved an exposure of 24524.3 kg days, i.e. 67.2 kg yr, taken over one year. Only single-scatters were considered, pulse shape discrimination techniques used, and a region of interest $3-11$~\keVee\ was chosen. 
Using Bayesian analysis, KIMS did not find any significant access of nuclear recoil events over background, with a total event rate below the DAMA annual modulation amplitude in the corresponding energy region. 
This was the first partially target independent test of the DAMA signal, as both experiments contained iodine in their crystals. 
However, the quenching factor of iodine is different in NaI(Tl) and CsI(Tl) resulting in a minor uncertainty as to which energy region in \keVee\ should be compared between these experiments \cite{Kim2010, Park2002}.

The KIMS team undertook a new run~\cite{Kims2014} from 2012 - 2014, adding two small NaI(Tl) crystals inside their detector, to more directly test the DAMA claim. The collaboration grew 6 R\&D NaI(Tl) crystals of total mass $\sim 45$ kg~\cite{Kims2016} with the Alpha Spectra Company, achieving a background level of $\sim 3$ counts \counts at $6$\keVee and a $<2$\keVee energy threshold thanks to removal of impurities in the powder through the growing phase along with strict background contamination controls.
Essentially the experimental run for KIMS-NaI was folded into COSINE, as discussed below, before significant limits could be placed on the DAMA claims.

\subsection{COSINE}
As a result of a collaboration between the DM-ICE and KIMS experimental teams, 106 kg of NaI(Tl) crystals that make up COSINE-100 have been operating since September 2016 in the Yangyang Underground Laboratory. 
The first 59.5 days of reported data, corresponding to a total exposure of 6303.9 \kgd, allowed highly constraining limits to be placed.
At 90\% confidence level, the WIMP-Na interaction cross-section for 10 \gevc WIMPs is reported at $1.14\times 10^{-40} {\,\rm cm^{2}}$~\cite{Cosine2018}.

Cosine-100 had an analysis energy threshold of 2 keV in six of the eight crystals with light yields of 15 photoelectrons per \keVee~\cite{Cosine100} while the remaining two crystals were removed from consideration due to higher thresholds and lower light yields. 

Each crystal was optically coupled to two PMTs and a `hit' was recorded when both measured at least one photoelectron within 200 ns. PMT-induced noise was removed by machine learning techniques, in particular boosted decision trees~\cite{friedman2001} applied to the pulse shape. 
The training dataset for the technique was provided during a two-week calibration campaign using a $^{60}$Co source. 

Active veto was also performed when both the liquid scintillator and crystal detected a multiple-hit. The combination of the veto and prior calibration of the pulse shapes permitted tagging of $^{40}$K generated 3 keV X-rays. 
Overall, the efficiency of the various methods to determine these background sources agreed to within $5\%$ which was then treated as a systematic error. 

Dark matter induced events were searched in single-hit energy spectra in the range 2 -- 20 keV for 18 WIMP masses~\cite{Cosine2018}, with no excess events attributable to a standard-halo WIMP detected. These 90\% confidence level exclusion limits are at a lower interaction cross-section than DAMA/LIBRA claim and further bring that result into question.

However the most recent publication~\cite{Cosine100b} from the COSINE collaboration using 1.7 years of data (exposure 97.7 \kgyr) hints at a potential signal. 
They report an annual modulation of amplitude $0.0092 \pm 0.0067$ \counts with a phase of $127.2 \pm 45.9$ days which is consistent, at 68.3\% C.L., with both the null hypothesis and DAMA 2 -- 6 \keVee best fit value~\cite{Cosine100b}. 

With a total of five years, i.e. $\sim 3$ years further, data exposure COSINE-100 expects to test the DAMA claims in a model independent way at $3\sigma$~\cite{Cosine100b}. 
Furthermore, a planned upgrade, COSINE-200, is already underway including the construction of a deeper underground site at Yemi Laboratory in Jeongseon County.

\subsection{ANAIS}
The Annual modulation with NaI(Tl) Scintillators (ANAIS) experiment \cite{Amare2017} is housed at the Canfranc Underground Facility in Spain, and is comprised of nine NaI(Tl) detector crystals each of mass $12.5$kg with an active veto system using a plastic scintillator.  This experiment follows in a lineage of annual modulation searches at Canfranc dating back to the nineties, one of the first such searches in the world~\cite{Sarsa1997}, which saw three $10.7$kg NaI scintillators capture two years of data (for a total exposure of $4613.6$\kgd; of which $1342.8$\kgd was used for annual modulation analysis).

ANAIS began~\cite{Cuesta2014} as a single NaI(Tl) bulk module, ANAIS-0, of $9.6$kg mass with two coupled PMTs contained in ultrapure copper enclosure. 
This experiment undertook a detailed event identification and selection study which informed the first run~\cite{Amare2014} of ANAIS-25 with two $12.5$kg NaI(Tl) crystals. 
Eacj crystal is surrounded by an OFE (Oxygen Free Electronic) copper cylinder, with a quartz windows at each end to optically couple Hamamatsu R12669SEL2 PMTs. ANAIS-25 saw the copper enclosed crystals placed within 10cm of archaeological lead, and further enclosed with 20cm of low activity lead, which itself resided in the PVC moderator box flushed with nitrogen~\cite{Amare2014} and an active muon veto using plastic scintillators across the top and sides of the entire experiment~\cite{Amare2018b}. 
ANAIS-25 focused on the determination of the potassium content of the crystals~\cite{Amare2014}, finding $^{40}$K levels of $1.25\pm0.11 {\rm mBq/kg}$ ($41.7\pm3.7$ ppb of potassium) which improved on ANAIS-0 by an order of magnitude, but not yet reaching the stated 20 ppb goal. 
A third crystal (D2) was then added to form ANAIS-37, a sufficient radiopurity was achieved that background rates of $<2$ counts \counts above 4\kevee could be expected for a 3x3 matrix of crystals (equivalent to $112.5$kg detector mass) in ANAIS-112~\cite{Amare2016}.

The experimental design for ANAIS resulted in a light collection yield of 15 photoelectrons per keV~\cite{Olivan2017,Coarasa2019}, which permitted a threshold energy for the detectors of 1 keV~\cite{Amare2018}. 
The background was measured~\cite{Amare2018b} to be $3.58\pm{0.02}$ \counts from the crystals produced by Alpha Spectra Inc. from Colorado, USA. 
This is a factor three higher than DAMA and will result in a 3$\sigma$~C.L test of a DAMA-like detection in the proposed 5 years of operation~\cite{anais2019,Coarasa2019}.

ANAIS is calibrated every two weeks using an external $^{109}$Cd source, exposed to the crystals through a Mylar window, allowing low energy calibration~\cite{Amare2018b}. 

The initial year and a half of data for ANAIS-112 has been recorded\cite{anais2019} resulting in 527 live days of exposure (reduced to 511 after vetoing of a seconds worth of data capture per muon trigger, equivalent to $157.55$ \kgyr). 
The reported data are consistent with the null hypothesis of no modulation\cite{anais2019}. 
This has been extended~\cite{Amare2019} to two years of data, equivalent to $220.69$ \kgyr exposure, and achieves an annual modulation best fit in the 2--6 \kevee range of $S_{\rm m} = -0.0029 \pm 0.0050$ \counts ($S_{\rm m} = -0.0036 \pm 0.0054$ \counts in 1--6 \kevee range) again, consistent with no modulation. Overall, ANAIS currently reports~\cite{Amare2019} a $2.6\sigma$ incompatibility with the DAMA result and is on track to achieve the aimed for $3\sigma$ test of DAMA within three years.

\subsection{SABRE}
The Sodium-iodide Active Background Rejection Experiment (SABRE) is the sole dual hemisphere effort capable of unambiguously detecting and determining the phase of a DAMA-like annual modulation signature~\cite{SABRE_SABREPOP}. 
Currently the Proof-of-Principle (SABRE-PoP) experiment is already housed~\cite{SABRE_MC} in Italy's LNGS while the Southern Hemisphere based-detector will be operated at the Stawell Underground Physics Laboratory in Victoria, Australia. 
This Southern detector in Australia is on track for commencement at the start of 2021. 

Each twin detector will ultimately comprise 50 kg of NaI(Tl) crystals, the ultra-pure crystals are grown in a technique developed by Princeton University that uses Sigma-Aldrich `Astro Grade' powder. 
These crystals~\cite{Xu2015} achieve potassium levels lower than $\sim 10 {\,\rm ppb}$ and levels or $^{238}$U and $^{232}$Th lower than $\sim 1 {\,\rm ppt}$. 
They are the purest NaI crystals produced to date and a marked improvement over DAMA, allowing target mass a factor five times smaller to constrain the latter in less exposure time.

Individual 5kg crystals are enclosed in a light tight copper cylinder, with two ultra-low background PMTs optically coupled to each end of the crystal. 
Bundles of these rods are then inserted inside the vessel with a bespoke glove-box, that is flushed with dry nitrogen to remove any radon or water vapour. 

The experiment employs Hamamatsu R11065-series PMTs tubes with a 35\% quantum efficiency which, when combined with an operating voltage of $\sim 1100 {\,\rm V}$ to reduce the afterglow noise, enables SABRE to test DAMA by reaching a threshold of $<1$ keV~\cite{Xu2015}.

SABRE was the first experiment that proposed the use of an active veto liquid scintillator system to suppress backgrounds. 
In principle, a $4\pi$ active volume around the crystal is employed to suppress the $^{40}$K background more efficiently. 
The fluid in the Northern Hemisphere-based SABRE-PoP is pseduocummene, while the SABRE-South is using a linear alkyl benzene solvent (primarily due to its higher flash point, and hence less volatile, nature). 
This proposal has been demonstrated to great effect in COSINE since.

A conservative estimate~\cite{SABRE_SABREPOP} of the background rate for SABRE is 0.4 counts \counts, but with both active veto and the current levels of ultra-pure crystals it is possible to reach of 0.1 counts \counts in the critical 2 -- 6 keV energy region~\cite{SABRE_SABREPOP}. 
Critically, even the conservative background can allow SABRE to refute or confirm DAMA at over $3\sigma$ confidence level in just three years exposure~\cite{Shields2015}.

\subsection{Other Direct Detection Technologies}
\label{sec:ExpTests_Other}
Beyond NaI crystals there exist a range of highly sensitive detector technologies using grammes of germanium to tonnes of xenon. 
The range of atomic weights and background noise levels provide a critically important breadth in sensitivity to candidate dark matter masses beyond the NaI crystal detectors described before. 

For example, semiconductor band gaps of $\mathcal{O}$(1 eV) enable microelectronics to be exquisitely sensitive to small ionisation signals from dark matter candidates as light as $\mathcal{O}$(100 \kevc )~\cite{Essig2016}. 
Whereas the ability of large, purified tanks of noble gases to achieve near zero backgrounds thanks to detection of both scintillation and ionisation signatures together~\cite{Davies1994} have meant that these facilities are often responsible for the most constraining of interaction cross-sections. 

\subsection{CoGeNT}

CoGeNT was designed to test the hypothesis that the DAMA modulation was produced by low-mass WIMPs (m$_\chi < 10$~\gevc). 
It uses a 440-g $p$-type Point Contact germanium (\ppcge) detector in the Soudan Underground Laboratory in Minnesota, USA. 
The advantage of a \ppcge is both a low threshold (0.5~keV) as well as very good pulse shape discrimination techniques to distinguish between signal and background \cite{Cogent2011}. 

In April 2013, based upon 1129 live days, the collaboration found a modulation in their event rate for low energy bulk events at an energy of $E=0.5-2$~\kevee\  with modest statistical significance of 2.2~$\sigma$. 
Depending on the details of their analysis, they find a period of $T=336\pm24$~days or $T=350\pm20$~days as well as a peak at $t_{\rm max} = 102\pm 47$~days compatible with DAMA's $t_{\rm max} = 136\pm 7$~days. 
Assuming the SHM, the signal could be contributed to a $m_\chi \approx 8$ \gevc WIMP \cite{Cogent2014}. 
They find a large fractional amplitude, which could entail a non-Maxwellian component for the local galactic halo \cite{Kelso2011}. 
In combination with other results including their planned upgrade C-4, with four detectors each approximately three times the mass of the original CoGeNT germanium diode \cite{Bonicalzi2013}, their data can help constraining potential models. 
However, an independent analysis of the data contributes the modulation to background in the bulk of the detector \cite{Davis2014}.

\subsection{CRESST}
The Cryogenic Rare Event Search with Superconducting Thermometers (CRESST), located at LNGS in Italy, measures light and heat output from CaWO$_4$ crystals to search for low-mass WIMPs. 
Advantages of their detector concept include a very low threshold as well as good discrimination power between electron and nuclear recoils due to the two readout channels. 
In 2012, the CRESST II collaboration published results from phase 1 based on 730~kg~days of data taken with eight detector modules showing an access in their acceptance region. 
None of the four major background sources investigated at that stage could sufficiently explain the data. 
WIMPs with one of two masses are compatible with the results: $m_\chi=25.3$~\gevc\ primarily from tungsten recoils with a significance of 4.7$\sigma$ or $m_\chi=11.6$~\gevc\ from oxygen and calcium recoils of roughly equal proportions with a significance of 4.2$\sigma$ \cite{Cresst2012}. 

In Phase 2, the collaboration investigated their excess with an improved setup primarily focussing on reducing the background. 
A low threshold analysis was performed with one of their modules called TUM40 (total mass of 249g), which showed overall the best performance in CRESST II Phase 2. 
This has an exceptionally low background event rate of 3.51 \counts, nearly an order of magnitude lower than commercially available CaWO$_4$ crystals~\cite{Cresst2015}. 
Based on 29.35~kg live days of non-blinded data and using the Yellin optimum interval method the lower mass WIMP with $m_\chi=11.6$~\gevc\ was clearly excluded and the higher mass WIMP with $m_\chi=25.3$~\gevc\ disfavored.\cite{Cresst2014}.

An even lower energy exploration phase~\cite{Cresst2016} was completed in August 2015, using the Lise module with the lowest energy threshold yet achieved by CRESST at $0.3$ \keVee for $52.2$\kgd exposure.
This module suffered a higher background than the previous Phase 2 run; 13 counts \counts across 1 -- 40 \keVee reduced to 8.5 \counts if the known $^{55}$Fe source events are excluded. The small target mass precluded significant limits being set on higher dark matter candidate masses but below $1.7$ \gevc the Lisa module set vastly more stringent tests than ever before, extending to $0.5$ \gevc for the first time.

CRESST-III explores even lower mass regions, having individual detector masses at 24g, an order of magnitude lower that allow even smaller heating events as well as scintillating light events from dark matter collisions to be recorded. 
In the first run completed in Feb 2018, CREST-III was able to achieve a nuclear recoil energy threshold of $30.1$ \eVnr, implying sensitivity to dark matter as low as $160$\mevc~\cite{Cresst2019}.

\subsection{CDEX}
At the China Jinping Underground Laboratory, in Sichuan, the China Dark Matter Experiment (CDEX~\cite{Cdex}) has undertaken several increasingly large target mass runs. 
Protected by a rock overburden of 2400m, the germanium detectors of CDEX-1 are enclosed in NaI(Tl) cylinders~\cite{Cdex2014} enabling competitive background suppression to very low energy thresholds ($0.32$ \keVee) for the Phase 1 run, the threshold reached was halved in the upgrade to CDEX-10~\cite{Cdex10} and will ultimately be $100$ \eVee for CDEX-1T~\cite{Cdex}.

Candidates were selected first on the timing correlation and basic pulse shape discrimination~\cite{Cdex2013} from the \ppcge. 
This first run had no active anti-Compton system, to allow the team to better measure the background level of the \ppcge itself~\cite{Cdex2013}. 
Only a passive shielding system was used in this first run, with consisting of 20 cm of oxygen-free high-conductivity copper surrounding the \ppcge, then a 20 cm layer of borated polyethylene, 20cm of lead and then a final 1 m layer of polyethylene.
The threshold was $400$ \eVee, an exposure of 16 \kgd for a fiducial target mass of 905 g, with at least 10 \counts background (increasing by a factor five at low 1 -- 1.5 \keVee energies due to $L$-shell X-ray events, well modelled by the observed higher energy $K$-shell transitions). 

Following the success of the first \ppcge tests at Jinping, the CDEX team then included an active anti-Compton background veto system by enclosing a germanium detector in a NaI(Tl) cylinder~\cite{Cdex2014}. 
The NaI(Tl) events are detected without delay, while there is a timelag within the germanium detector as the charge drifts, this time-difference then allowed (anti-) coincidence events to be selected. Indeed, the anti-coincidence of the AC effectively removed all of the $\gamma$-ray induced backgrounds~\cite{Cdex2017}. 
Using an active target mass of 0.919 kg with a greater exposure of $53.9$ \kgd and a reduction of background rates to 4.09 (4.22) counts \counts at 0.475 -- 0.575 (1.975 -- 2.075) \keVee this run~\cite{Cdex2014} was sufficient to rule out DAMA masses at $6$ and $20$ \gevc.

A lower threshold 0.939 kg germanium detector was undertaken as part of CDEX-1B, with 17\% efficiencies reported at a greatly reduced $160$ \eVee threshold (as compared with 80\% at 475 \eVee previously)~\cite{Cdex2017}. 
This run undertook an annual modulation search with an exposure of 737.1 \kgd and reported background rates, after removal of $L$/$M$-shell X-rays, of $\sim 7\,(2.5)$ \counts at energies of 0.16 (2.5) \keVee~\cite{Cdex2017}. 
This effectively surpassed the CRESST-II limits~\cite{Cresst2016} demonstrating the maturity of the electronics / noise for reduced thresholds as well as suppression of background contamination at CDEX. 
Further analysis~\cite{Cdex2019} treated phase of an annual modulation as a free parameter across 3.2 years of live data (out of the 4.2 year span) and, for WIMP masses between 3.2 -- 17 \gevc, found to be consistent within $1\sigma$ of the null hypothesis of no detection across $0 - 2\pi$ values in the phase. 
The Migdal effect was incorporated to further decrease the mass window by an order of magnitude~\cite{Cdex2019b}, setting spin-independent mass limits of $3 \times 10^{-32} -- 9\times 10^{-38}$ \cms for masses between $75$\mevc -- $3$\gevc using the annual modulation null detection.

Following the success of CDEX-1, an order of magnitude greater detector mass was constructed to form CDEX-10~\cite{Cdex10} with the 10 kg of \ppcge detectors now distributed in three triple-element `strings' immersed in liquid nitrogen.
This $LN_{2}$ provides super shielding and better control of backgrounds than the previous reliance on high-atomic number materials~\cite{Cdex10}. 
Unfortunately, due to one string being upgraded, and others suffering faulty cabling and high backgrounds only one of the nine detectors overall were reported in the latest CDEX-10 run~\cite{Cdex10}. 
However in the low energy range the detector reported corrected bulk background counts of 2.47 (2.15) \counts in the energy range 0.16 -- 0.26 (1.96 -- 2.06) \keVee, in an exposure of 102.8 \kgd of fiducial mass 0.939 kg over an energy range 160 \eVee to 12 \keVee~\cite{Cdex10}. 

Ultimately the upgrade path will continue to CDEX-1T, i.e. a one tonne active mass experiment, with 1 \tonyr exposure, threshold of 100 \eVee and background of 0.01 \counts will achieve $10^{-43}$\cms for WIMP masses 2 -- 10 \gevc, and orders of magnitude more  constraining for masses below 5 \gevc than any other detector, essentially reaching the Solar neutrino floor~\cite{Cheng2017}.

\subsection{DAMIC}
The possibility of utilising charge-coupled devices (CCDs) for low mass dark matter detection is a relatively recent technology path~\cite{Chavarria2016} and its experimental use has been demonstrated in DAMIC (dark matter in CCDs) at SNOLAB~\cite{DAMIC}. 
The low read out noise of modern CCDs, as well as the relatively low mass of silicon as an atomic target make DAMIC most sensitive to 1 -- 20 \gevc candidates. 

The development of the CCDs by Lawrence Berkeley National Laboratory MicroSystems Lab~\cite{DAMIC_CCD} was in fact initiated for the Dark Energy Survey camera~\cite{DECam}. Ionization charge from a collision within a $15 \times 15$ \um$^{2}$ pixel drifts along the electric field in the z-axis, with the holes (i.e. charge carriers) collected and held near the p-n junction, less than 1\um below the gates~\cite{DAMIC_CCD}. 
This charge is held at the gates throughout hour- to day-long image exposures until the device is read out~\cite{DAMIC}. As noted in~\cite{DAMIC} the CCD's low dark current ($<10^{-3} {\rm e}^{-} {\rm [pix day]}^{-1}$) allow several day exposures with minimal addition of noise.

Placing two 8Mpix CCDs inside a copper vacuum vessel cooled to 120K, with an inner lead shield to block $\gamma$ radiation sourced from an ancient Spanish galleon,~\cite{DAMIC} achieved 0.6 \kgd exposure in a demonstration run. This measurement reached a threshold of 60 \eVee with a background of 30 \counts, although this decreased over the run as radio purity of the setup was improved~\cite{DAMIC}.

Ultimately eighteen 16 Mpix CCDs (5.8 g each) will be used to form the upgraded detector DAMIC100, containing 100g of bulk silicon~\cite{DAMIC100}.

\subsection{CDMS}
\label{sec:CDMS}
The Cryogenic Dark Matter Search (CDMS-II) Collaboration experiment, located in the Soudan Mine in Minnesota, USA, uses germanium and silicon detectors at low temperatures of $\lesssim 50$~mK in order to detect ionization and athermal phonons (``heat'') generated by WIMP induced nuclear recoils. 
Similar to CRESST, the readout of two different channels allows discrimination between electron and nuclear recoils. 
However, this is less effective at low recoil energies of $\lesssim10$~\keVnr\ as expected from interaction with WIMPs with masses $\lesssim 10$~\gevc, because background events start populating the signal region and the ionization signal is comparable to readout noise.  

To specifically test the signal region from DAMA ($m_\chi\approx 100$~\gevc), the collaboration~\cite{Cdms2010} used only the 19 germanium towers due to their greater sensitivity to spin-independent WIMP scattering, with the silicon used to veto backgrounds. 
The exposure time, after excluding poor detector performance, was 612 \kgd in which a blind analysis was performed. 
After unblinding, two WIMP-like events (at 12.3 and 15.5 \kevee) were revealed but the probability of two or more background events in this run was 23\%~\cite{Cdms2010} and hence cannot be interpreted as significant. 
This run all but ruled out the high mass ($\sim 100$\gevc) region of the DAMA claimed detection. 

Follow up studies focussing on the low energy / mass region used only the eight germanium towers with the lowest threshold (2 \keVee), and again the remaining Si / Ge detectors were used to veto backgrounds~\cite{Cdms2011}. 
The germanium detector mass for this reduced experiment was 1.912 kg out of a total of 4.6kg, providing 241 \kgd of raw exposure. 
Using the Yellin optimum interval method together with the SHM, the interpretation of DAMA's signal as spin-independent elastic scattering of low mass WIMPs could be excluded.

Similarly, these results are incompatible with CoGeNTs claims, for which the same target material was used \cite{Cdms2011}. 
It has later been suggested that only a small fraction of the low-energy excess events in CoGeNT are due to WIMPs, making all results consistent \cite{Hooper2012}. 
This scenario was tested by CDMS in a study specifically focusing on a potential annual modulation signal. 
Data was collected over nearly two years and again only the data from the eight germanium detectors with the lowest threshold was used to search for WIMPs, while other detectors were used as active veto. 
No evidence for an annual modulation was found, constraining the magnitude of any modulation to $< 0.06$~event [\keVnr~kg~day]$^{-1}$ in the $5-11.9$~\keVnr\ range at 99\% confidence level (C.L.). 
This disfavors the interpretation of CoGeNT data due to WIMP scattering  at $>98$\% \cite{Cdms2012}.

The publicly available CDMS data was reanalysed~\cite{Collar2012} focussing on the low-energy region using an unbinned maximum likelihood analysis. In this work they found a 5.7$\sigma$ C.L. significance for an exponential excess. 
However, the CDMS collaboration analyzed the data as well and specified that the excess is observed not just in single- but also multiple-scatter events, implying it is not caused by WIMPs but rather some background, which was not properly included before \cite{Cdms2015}.

A separate analysis was undertaken using the silicon detectors alone~\cite{Cdms2013}, in which 3 of the 11 silicon detectors were found to be problematic (wiring failures / instability phonon response) resulting in 8 stacks for a total mass 0.848 kg in Si over the period July 2007 -- Sept 2008. 
No evidence for a WIMP signal was found based on an exposure of 140.2 \kgd, although after applying selection criteria the analysis was equivalent to 23.4 \kgd for the recoil range 7 - 100 \keVee for a WIMP of mass 10 \gevc \cite{Cdms2013}. 
A blind analysis of this silicon data did revealed three events potentially from WIMP interactions. 
The probability of a background source producing three or more events in the signal region was given with 5.4\%, while the highest likelihood occurs for a WIMP with a mass of 8.6~\gevc\ \cite{Cdms2013}. 
The claim was not yet tested with more sensitive silicon data as the CDMS-II upgrade, SuperCDMS, focused on the germanium technology path.

\subsection{SuperCDMS}
\label{sec:supercdms}
Focusing on the low energy regime, the CDMS team upgraded the germanium detector technologies, to create SuperCDMS \cite{Supercdms2014} with $15 \times 0.6$ kg germanium crystal stacks grouped into five towers. 
Further improving on the previous CDMS technique of ionisation and phonon sensors by now interleaving them between stacks, the discrimination between nuclear and electron recoils further reduces the background rate for SuperCDMS to an order of magnitude lower than CDMS-II~\cite{Supercdms2014}.

A limited SuperCDMS low-ionization threshold experiment (CDMSlite) was undertaken using a single 0.6 kg germanium detector from SuperCDMS~\cite{CDMSliteI} with the second, larger run in 2015~\cite{CDMSliteII} that achieved 70.1 \kgd exposure. 
Voltage-assisted Luke-Neganov amplification of the ionisation energy from interacting particles was employed with a larger bias voltage (-70V to 0) than previous CDMS runs~\cite{CDMSliteI}. 
This enhanced the phonon signal and permitted electron recoil thresholds as low as 56 \eVee~\cite{CDMSliteII}. The background rate for the lowest energy regime, 0.056 - 0.14 \keVee were 16.33 \counts but this reduced to $\approx 1$ \counts for 0.2 - 1.0 \keVee overall~\cite{CDMSliteII}.
This run excluded new regions of WIMP parameter space between 1.6 and 5.5 \gevc and demonstrated the value of including both higher bias and phonon resolution in the SuperCDMS-SNOLAB upgrade.

The full mass range SuperCDMS experiment~\cite{Supercdms2018} detected a single candidate event after an exposure of 1690~\kgd consistent with backgrounds. 
This measurement has set new limits for DM-germanium interactions in the mass range 13 -- 127 \gevc~\cite{Supercdms2018}. 
These limits will not be surpassed until the establishment SuperCDMS in its new home - SNOLAB - in 2020 which will set constraints for low mass (0.5 -- 10 \gevc) dark matter candidates to within an order of magnitude of the neutrino floor where coherent scattering of Solar neutrinos is the limiting background~\cite{SupercdmsSNOLAB}.

\subsection{EDELWEISS}
Based in the Modane Underground Laboratory in France, under 1800m of rock, EDELWEISS (Expérience pour DEtecter Les WIMPs En Site Souterrain~\cite{EdelweissI}) utilises germanium-based bolometers to measure both phonon and ionisation signals at cryogenic temperatures, similar to CDMS (Sec~\ref{sec:CDMS}). 
The first run was completed in 2003, and consisted of three $0.32$ kg Ge detectors operated at 20mK by dilution fridges, reaching a total fiducial exposure of 62 \kgd and an energy threshold of 13 \keVee during four months of operation~\cite{EdelweissI}.

The measurement of heat deposition is by the thermal phonon sensors, coupled neutron transmutation doped germanium thermometric sensors on each detector~\cite{EdelweissI}, with ionisation measurement by two aluminium electrodes operating at a voltage bias of $-4.0$ V for all but one of the first phase runs~\cite{EdelweissI}. 

For EDELWEISS-II~\cite{EdelweissII}  the number of Ge detectors was increased to ten, for a total mass of $\sim 4 kg$ (but only $1.6 kg$ fiducial~\cite{EdelweissII}) and a larger run of 14 months operation, with 384 \kgd exposure. 
Thresholds for nuclear recoil events in the analysis was conservatively set at 20 \keVnr in a search optimised for high mass, $>50$ \gevc, WIMP candidates~\cite{EdelweissII}.
For high mass 85 \gevc candidates the WIMP-nucleon cross-section limit at 90\% C.L. in this analysis~\cite{EdelweissII} was $4.4 \times 10^{-44}$ \cms, based on five nuclear events above 20 \keVnr whereas 3 would be expected from background.

A low-mass analysis was undertaken with a subset of the 4 EDELWEISS-II~\cite{EdelweissIILowM} detectors, in a similar manner to CDMS~\cite{Cdms2011} as discussed in Section~\ref{sec:CDMS}. 
This low-mass EDELWEISS effort searched for nuclear recoils exclusively below 20 \keVnr (a threshold limit of 5 \keVnr was set) in a total exposure of 113 \kgd and in the search region of interest only a single background event was detected~\cite{EdelweissIILowM}. 
For a 10 \gevc mass candidate this implied~\cite{EdelweissIILowM} the WIMP-nucleon spin-independent cross section at 90\% C.L. of $10^{-41}$\cms.

EDELWEISS-III~\cite{EdelweissIII} is the latest completed run by the collaboration which employed both larger detectors (820 -- 890 g) and an increased numbers of such detectors (36 in total although only 24 were used~\cite{EdelweissIIIb}), now equipped as Fully InterDigitized electrodes (FIDs) for rejection of near-surface events~\cite{EdelweissFID} allowing confirmed detection of events from the bulk of the fiducial detector volume. 
Additionally the entire crysostat containing the experiment is housed inside both lead and polyethelene shielding as well as an active muon veto system based around a plastic scintillator (Bicron BC-412) of total surface area 100 $m^2$ split above and below the cryostat~\cite{EdelweissIIIb}.

The 10 month run resulted in a 582 \kgd exposure of the array of bolometers and, for the low mass run of 4 -- 30 \gevc candidates, focused on the 2.5 -- 20 \keVnr events~\cite{EdelweissIII}. 
The expected background rate of 6 events in the entire run (an equivalent of $6 \times 10^{-4}$ \counts) and could set 90\% C.L. cross-section limits for 5 (20) \gevc of $4.3 \times 19^{-40}$ ($9.4 \times 10^{-44}$) \cms \cite{EdelweissIII}.

The EDELWEISS collaboration have since undertaken EDELWEISS-Surf~\cite{EdelweissSurf}, an above ground measurement experiment for low $<1$ \gevc masses using a 10mK dry dilution cryostat at the Institut de Physique Nucl{\'{e}}aire de Lyon based around a 33.4 g germanium detector with a neutron-transmuation doped germanium sensor similar to previous detectors to date. 
A benefit of being on the surface is that strongly interacting dark matter models can be tested due to the large differences in these models with the asymmetric Earth-shielding effects when on the surface~\cite{EdelweissSurf}.

The next-generation experiment for EDELWEISS is a joint effort with the team from CRESST and ROSEBUD to form EURECA (European Underground Rare Event Calorimeter Array) which will create the infrastructure for up to 1000kg cryogenic experiments on the future~\cite{EURECA}. 
Initially, the collaboration will adopt the current Ge and CaWO$_{4}$ detectors, based on their respective technology paths but may also include the specific germanium detectors from SuperCDMS (described in Section~\ref{sec:supercdms}). 
The first stage will see this infrastructure and 150 kg of detectors deployed, the second stage will proceed with an additional 850 kg (or a yet to be decided ratio of technologies) to reach the target tonne mass~\cite{EURECA}. 
This initial phase is already competitive with target cross-section interaction limits of $3 \times 10^{-46}$ \cms and an ultimate sensitivity of $2 \times 10^{-47}$ \cms based on residual backgrounds in the region of interest of $10^{-2}$ and $<10^{3}$ [kg~y]$^{-1}$ respectively~\cite{EURECA}.

\subsection{XENON}

XENON is a series of dual-phase time projection chamber experiments with growing target mass placed at LNGS in Italy. 
Due to the two phases (liquid and gas) of xenon, they also receive two signals, a prompt scintillation signal and a delayed ionization signal, which can be used to distinguish electron and nuclear recoil. 
They also receive position information, which helps to further identify background events. 
Neither XENON10, XENON100, nor XENON1t has so far found an excess of events above their expected background \cite{Xenon2007, Xenon2011, Xenon2012, Xenon20171t}, excluding the standard WIMP interpretation in their respective parameter space. 

The XENON collaboration investigated a few alternative explanations for the DAMA signal. 
In 2015, they published a study based on around 225 live days from XENON100 to probe if the DAMA signal could be due to certain types of leptophilic dark matter suggested e.g. by \cite{Kopp2009}. 
To be conservative they used the most challenging to exclude case of a fully modulating dark matter signal, thus assume no constant dark matter contribution. 
A 70 live days window around the peak time of the modulation at 2 June was used to search for an access in their event rate above their background. Events were required to be single scatters in the fiducial volume with prompt and delayed signal in the correct energy range.
In the first analysis, axial-vector couplings between dark matter and leptons were assumed. 
This case is independent of the assumed halo model since xenon atoms and iodine anions have very similar electron structure and thus nearly identical momentum-space wave functions while the contribution of sodium are two orders of magnitude smaller. 
No excess above background was found, excluding this model as explanation for DAMA's modulation signal at $4.4\sigma$ confidence level.
Next was considered a kinematically mixed mirror dark matter in which DAMA's signal was explained as dark matter-electron scattering \cite{Foot2014}. To be able to compare both experiments, a constant scaling factor using the number of loosely bound electrons and target atoms in both cases has to be applied. 
Again, no excess was found, excluding this the DAMA signal at $3.6\sigma$ confidence.
Finally, luminous dark matter with a 3.3\kevc mass splitting between states connected by a magnetic dipole moment operator was investigated \cite{Feldstein2010}. 
The corresponding signature is independent of the target material but again, no excess above background was found, excluding this as DAMA's signal with $4.6\sigma$ confidence limits \cite{Xenon2015}.

Using the same data, they also studied the possibility of magnetic inelastic dark matter (MiDM) as potential explanation for DAMA's modulation signal. 
This type of dark matter is primarily motivated by the fact, that iodine is distinguished from other target materials due to its large mass as well as large magnetic moment, and MiDM would thus produce a significantly higher event rate in iodine targets then others \cite{Chang2010}. 
The MiDM particle would first inelastically scatter off the target nucleus, which later de-excites, creating a unique signature in the detector. 
No such event has been found, excluding WIMPs with masses of 58~\gevc\ and 123~\gevc\ as proposed to explain the DAMA signal with $3.3\sigma$ and $9.3\sigma$, respectively \cite{Xenon2017mag}.

Another study \cite{Xenon2017mod} including two more data sets resulted in a total of 477 live days spanning nearly 4~years that was searched for electron recoil event rate modulations.
Single interactions in the energy region between 2.0--5.8~keV range were investigated as potential signal and the 5.8--10.4~keV region was used as control band similar to DAMA's analysis. 
Quality cuts were performed and potential correlations between variations in detector parameters (found to be less than 2\%), or background and signal variation studied. 
Total single background events for XENON100 were $\approx 1$ per day in the low energy region of interest, resulting in a rate of $2.6 \times 10^{-3}$\counts.  
An unbinned profile likelihood analysis was used to determine the statistical significance of a potential modulation signal, which was found to be $1.9\sigma$. 
However, after fixing the modulation period to one year the resulting DAMA modulation amplitude is far larger than that observed by XENON100 and excluded at $5.7\sigma$~\cite{Xenon2017mod}. 

The latest Xenon1T searches for light dark matter~\cite{Xenon2019a} have further reduced backgrounds, achieving $<10^{-3}$\counts above a threshold of 400 \eVee, and after 22 \tond have set new exclusion limits for WIMP candidates in the range 3 -- 6 \gevc~\cite{Xenon2019a}. 

A separate analysis that searches for secondary emissions following a collision has opened up a new mass regime as low as 60 \mevc~\cite{Xenon2019b}. 
This is because the recoiling nucleus from an elastic WIMP collision exhibits a momentum change relative to the initial orbital electrons causing a polarisation of the recoiling atom and kinematic boost of the electrons~\cite{Xenon2019b}. 
The search strategy can then be~\cite{Xenon2019b} to detect either resulting Bremsstrahlung emission~\cite{Kouvaris2017} from the depolarisation of the xenon atom, or secondary radiation signal from the ionisation / excitation of the atom in the kinematic boost of the electron known as the Migdal effect~\cite{Migdal1941, Ibe2018} as described in Section~\ref{sec:ER_Detector}. 
This electronic recoil search extended the mass exclusion range of dark matter from XENON1T down to 1.8 \gevc but with the possibility that XENONnT cross-section limits could be extended to masses as small as 60 \mevc~\cite{Xenon2019b}.

The ultimate planned upgrade path~\cite{Xenon2016} will see an order of magnitude improvement on cross-section limits from XENON1T as XENONnT is deployed with 8 (6 fiducial) tonnes of xenon which, after a 20 \tonyr exposure, will constrain the WIMP-nucleon cross section to $10^{−48}$\cms at 50 \gevc~\cite{Xenon2016}.

\subsection{ZEPLIN}
In the UK's Boulby Underground Laboratory a series of staged liquid xenon experiments known as ZEPLIN, ZonEd Proportional scintillation in LIquid Noble gases, with ZEPLIN-I undertaken in 2005~\cite{ZeplinI}, ZEPLIN-II in 2007~\cite{ZeplinII} and two runs of ZEPLIN-III culminating in 2011~\cite{ZeplinIIIa,ZeplinIIIb}. 
ZEPLIN-II~\cite{ZeplinII} pioneered the dual signal approach to discriminating nuclear / electron recoils as used in LUX for example, with ZEPLIN-III~\cite{ZeplinIIIa} containing an active mass of $\sim 12$ kg liquid xenon monitored by 31 PMTs for both prompt scintillation signals in liquid and delayed electroluminscence signal from the gas phase above it. 
The electron recoil background was dominated by these PMTs with 10.5 \counts from these alone, and essentially a further additional count from everything else. 
The dual signal approach allows almost complete removal of these backgrounds from confusion with nuclear, i.e. WIMP, collision~\cite{ZeplinIIIa}.

The fiducial volume after removal of backgrounds was defined as 6.5 kg of xenon~\cite{ZeplinIIIa} and, although the first science run achieved 847 \kgd exposure, after cuts to focus on the WIMP energy search range of 2 -- 16 \keVee this exposure was reduced to 127.8 \kgd~\cite{ZeplinIIIa}. 
The limits for this run within a factor two of the spin-independent cross-section constraints of CDMS-II~\cite{Cdms2009} across the mass range 10 -- 1000 \gevc. 

The second science run of ZEPLIN~\cite{ZeplinIIIb} replaced the PMT setup that reduced the PMT-originating $\gamma$-ray activity 40-fold, at the expense of dramatically poor optics, as well as upgraded the anti-coincidence veto system. 
This involved a tonne of plastic scintillator surrounding the existing ZEPLIN-III target and observed with PMTs~\cite{ZeplinIII_AC}.

The fiducial exposure was 1344 \kgd for this run, but after cuts, the net effective exposure for a 50 \gevc candidate WIMP is 251 \kgd~\cite{ZeplinIIIb} in the range 7.4 -- 29 \keVnr with an almost background-free experiment of just 8 events in the entire run corresponding to $< 3 \times 10^{-4}$\counts as a background. 
The future upgrade for path is a joint experiment with LUX, as described below, known as LUX-ZEPLIN~\cite{LZ}. 

\subsection{LUX}
Established in Lead, South Dakota, at the Sanford Underground Research Facility, the Large Underground Xenon (LUX) detector has established amongst the most constraining limits yet\cite{Lux2017}. 
LUX comprises of 370 kg of liquid xenon, of which 
250 kg is actively monitored in a dual-phase (liquid-gas) time-projection chamber (TPC)~\cite{Lux2014}. 
External shields of 300 tonnes of water, and 20 tonne of steel, suppress background $\gamma$-ray events completely with an overall background rate of order $1.7 \times 10^{-3}$ \counts expected~\cite{Lux_rad}.

Prompt scintillation events within the liquid xenon can also produce ionisation electrons that then drift into the gas phase under electric field, producing electroluminescence~\cite{Lux_elum}. 
The two light signal events are detected by two arrays of 61 PMTs to accurately reconstruct the deposited energy in the collision as well as use their ratio to discriminate between nuclear and electron recoils~\cite{Lux2014}.

As defined in \cite{Lux2014} the expected elastic scattering signal of a WIMP-like collision is a single prompt scintillation event, greater than 0.25 photoelectrons registered by at least two PMTs within a window of 100ns all with a corresponding electroluminescence in the gas phase no later than the maximum drift time of 342 \us (the latter light signal  must also be greater than 200 phe, equivalent to 8 extracted electrons). 

The equivalent threshold for detection is set at 3 \keVnr but LUX has sensitivity down to 0.7 \keVnr albeit at reduced efficiency~\cite{Lux2015}. 
The current largest run of LUX achieved exposure of $3.35 \times 10^{4}$ \kgd~\cite{Lux2017} which, when combined with previous runs, ruled out interaction strengths greater than $1.1\times10^{-46}$\cms at 90\% CI for WIMP masses at 50 \gevc.
 
While LUX has now been decommissioned, a joint collaboration with the ZEPLIN team to build a next generation detector, the 7 ton LUX-ZEPLIN~\cite{LZ} is on track to begin operations in 2020 at Sanford~\cite{LZ_TDR} to improve the interaction strengths limits by almost two orders of magnitude to $3\times 10^{-48}$\cms for similar WIMP masses.

\subsection{PandaX}
\cite{PandaXII}
The deepest currently operating detector, the Particle and Astrophysical Xenon Detector (PandaX~\cite{PandaX1a}) is located in the China Jinping Underground Laboratory. 
PandaX is a large-scale dual phase xenon detector, employing a Zeplin-like dual signal discrimination technique from the gas and light phases to differentiate electronic and nuclear recoils~\cite{PandaX1b}.
The latest observational run of PandaXII~\cite{PandaXII} had total (fiducial) mass of Xenon of 580 (330 -- 360) kg, approximately twice the target mass of LUX. 
At the energy range of 1 -- 10 \keVee~\cite{PandaXbkg} with a 54 \tond exposure, and background\footnote{improved neutron background estimates~\cite{PandaXbkg} reduced this overall background to $0.66 - 0.47 \times 10^{-3}$\counts for the two separate runs that made up the full exposure} of $0.8 \times 10^{-3}$\counts~\cite{PandaXII} it essentially matched the constraining power of the most recent 2017 LUX measurements~\cite{Lux2017}, as shown in Fig.~\ref{fig:dmlimits}. 

PandaX is currently being upgraded to PandaX-4T, a four ton experiment with total mass of 2.8 ton in the fiducial volume~\cite{PandaX4T}. 
In the critical range of 1–-10 \keVee, the total electron recoil background $4-9\times 10^{-5}$ \counts and a nuclear recoil background of $2-8 \times 10^{-7}$ \counts~\cite{PandaX4T}. 
It is predicted that a 5.6 \tonyr exposure will see PandaX-4T constrain spin-independent dark matter - nucleon $\sigma = 6 \times 10^{-48}$\cms for 40 \gevc mass WIMP.

\subsection{XMASS}
XMASS is a single phase liquid xenon detector with a total mass of 835 kg, located at Kamioka Observatory in Japan at a depth of 2700 m water depth equivalent~\cite{Xmass1}. 
The xenon is contained within a seal chamber formed as a pentakis-dodecahedral of triangular copper plates each with surface mounted PMTs (642 in total), and within this active volume is 832 kg of liquid xenon~\cite{Xmass2019}. 
The inner detector is immersed within a 10m diameter copper cylinder-vessel of purified water and monitored by 72 20-inch PMTs as an active veto water-Cherenkov outer detector. 
After cuts, XMASS claimed a background rate of 1.17 (0.028) \counts at 1.1 (5.0) \keVee~\cite{Xmass2016}.

Events were required to trigger 4 or more PMTs in a 200~ns coincidence window without simultaneously triggering the water veto. 
Further quality cuts were applied and 1.1--15~\keVee\ chosen as region of interest. 
Regular calibration showed a variation of $\sim \pm2.5$\% in photoelectron yield depending on position and time after corrections. 

The first major run spanned 359.2 live days between November 2013 and March 2015~\cite{Xmass2016}, and the data was binned into roughly 10 live days each and then 0.5~\keVee\ energy bins to perform a least-square fit. 
Neither an analysis explicitly assuming a WIMP interaction nor a second one independent of the specific dark matter modulation found any indication of a modulation of the event rate after nearly 300 \tond of exposure~\cite{Xmass2016}. 
XMASS is unable to distinguish between electron and nuclear recoils, and hence could not rule out that the possibility that either electron or gamma-ray events produced by dark matter could explain the DAMA results~\cite{Xmass2016}. 
This analysis was extended~\cite{Xmass2018} to 800 live days of the same target, with $1.82$ \tonyr exposure and an energy threshold of 1 \keVee (corresponding to $4.8$ \keVnr), and a background rate after cuts of $\sim 0.75$ \counts at $1.0$\keVee. 
The greater exposure time resulted in the statistically anomaly of `negative' residual modulations reported previously~\cite{Xmass2016} disappearing and further improved the constraints on the WIMP-nucleon cross-section by a factor of 2~\cite{Xmass2018}. 
This effort found no statistically significant periodicity (with period 50 -- 600 days) in the amplitude between energies 1 to 6 \keVee~\cite{Xmass2018}. 

An additional analysis of 705.9 live days, in a reduced fiducial volume of 97 kg of liquid xenon that resulted in a greatly reduced background rate of $(4.2\pm 0.2) \times 10^{-3}$ \counts at 5\keVee, and a threshold of 2 \keVee, has provided one of the most stringent limits on the WIMP-nucleon cross-section for a xenon detector to date; no greater $2.2 \times 10^{-44}$\cms for 60\gevc at 90\% C.L. based on the 68.5 \tond exposure~\cite{Xmass2019}.

The XMASS collaboration ultimately are aiming to deploy XMASS-II, a 20 tonne liquid xenon detector, but an intermediate step with 5 tonnes, XMASS-1.5 is currently under construction with numerous design improvements including lower background PMTs with greatly increased light collection especially of emission traveling parallel to the inner detector surface~\cite{Xmass_1o5}.

\subsection{DarkSide}
Beginning in Italy's LNGS facility, the DarkSide collaboration~\cite{DarkSide15} have undertaken several increasingly large liquid argon TPCs. 
The basic design is three nested detectors; the argon TPC, immersed within a 4m diameter spherical vessel containing a liquid scintillator active veto, all within a a larger water Chernkov detector~\cite{DarkSide15}.

The water tank is a 11m diameter and 10m high cylinder, originally from Borexino Counting Test Facility, lined with laminated Tyvek-polyethylene-Tyvek reflector and filled with 1000 tonnes of high purity water~\cite{DarkSide15}. 
Cherenkov photons produced by relativistic particles, especially muons, are captured by 80 PMTs. 
The active veto consists of 30 tonnes of borated liquid scintillator (equal parts pseudocummene and  trimethyl borate) with $2.5g/l$ of a wavelength shifter Diphenyloxazole. 
The inner surface is laminated with  stable reflector foils of Lumirror and scintillation photons within this vessel are detected by 110 PMTs.

Finally, the liquid argon TPC is similar in concept to the previously described xenon experiments (e.g. PandaX, Lux or XENON) with a prompt scintillation signal from collisions, with the light-curve of the pulse shape discriminating between nuclear and electron recoils. 
The resulting electrons that drift in the electric field cause a secondary scintillation signals as they accelerate, in direct proportion to the ionisation strength. 
The combination of these two events allows both 3D reconstruction of the energy deposition, as well as further discrimination of backgrounds~\cite{DarkSide15}.

The initial run with total exposure of 1442 \kgd used atmospheric argon, which is contaminated by cosmic-ray produced radioactive $^{39}$Ar~\cite{DarkSide15}. This background was removed in the first main science run for DarkSide-50~\cite{DarkSide16} by using ultra-low background argon. 
This argon is sourced from underground gas wells that contain naturally decayed levels of $^{39}$Ar~\cite{UAr1} which was then cryogenically distilled to further decrease the levels of $^{39}$Ar~\cite{UAr2,UAr3,UAr4}, ultimately reducing background by a factor of $10^3$ to atmospheric argon.

The 46.4 kg of target argon in DarkSide-50~\cite{DarkSide16} produced a background-free null detection with exposure of 2616 \kgd that, when combined with the previous run~\cite{DarkSide15}, resulted in 90\% confidence upper limits for spin-independent cross-sections of $2\times 10^{-44}$\cms for 100\gevc WIMP masses.

Due to the relatively low atomic mass of argon compared with other noble gas experiments, DarkSide-50 could competitively constrain WIMP masses below 20\gevc with a relatively small target mass. With a threshold of 0.1\keVee the low-mass search~\cite{DarkSide18a} undertook a 6786 \kgd exposure and set new limits on WIMP interactions in the mass range 1.8 -- 6 \gevc, although further reduction of the threshold to just 3 electron events (equivalent to 0.05 \keVee) and considerations of `heavy' mediators (as explained in Section~\ref{sec:ER_Particle}) had DarkSide-50~\cite{DarkSide18a} upper limits surpass even the constraints of XENON10 and XENON100 for 30 -- 100 \mevc WIMP masses.

Ultimately the collaboration (in partnership with members of DEAP, discussed below) are aiming to build a 23 (20) tonne (fiducial) mass liquid argon time projection chamber known as DarkSide-20k~\cite{DarkSide20k}. 
Thanks to a rejection factor discriminating between electron and nuclear recoils of greater than a billion, and additional veto and silicon photomultipliers tracking events within the TPC, the collaboration believe~\cite{DarkSide20k} a background free experimental run of 100 \tonyr exposure is possible that would constrain WIMP-nucleon cross-sections for 1 \tevc candidates to $1.2 \times 10^{-47}$\cms.

\subsection{DEAP-3600}
The DEAP (Dark matter Experiment using Argon Pulse-shape discrimination) collaboration is based in SNOLAB where they have deployed two liquid argon detectors (the first generation DEAP-1~\cite{Deap1} and the currently operating DEAP-3600~\cite{Deap3600}). 
Although utilising {\it single}-phase liquid argon as a target, DEAP-3600 was similar in design to DarkSide's central spherical vessel monitored by a number of PMTs all submerged within a large Cherenkov water tank~\cite{DeapDesign}. 
Of novel distinction with DarkSide was also the employment by the DEAP collaboration of a unique acrylic vessel design for DEAP-3600 with lightguides directing scintillation emission to 255 8-inch Hamamatsu R5912-HQE PMTs~\cite{Deap1_bkg}. 

The initial prototype, DEAP-1, ran between 2007 and 2011 using 7 kg of liquid argon in a cylindrical acrylic housing with windows to two PMTs and wavelength shifter 1,1,4,4-tetraphenyl-1,3-butadiene (TPB) which shifts the liquid argon scintillation light to a peak wavelength of 440 nm~\cite{Deap1,ArShifter}. 
Notable enhancements through this proof of concept phase include upgrades to the PMTs, measurements of radon daughters contamination of the active volume surfaces and improvements in the pulse shape discrimination to reduce misidentification of electromagnetic events as nuclear recoils~\cite{Deap2016}. 
Indeed, a fundamental advantage for argon as a target is the emission of the scintillation light from a two dimmer states of vastly different lifetimes, the singlet is just 6ns while the triplet state is $\sim 1.5$\us, with a different population of the states from lower-energy electron versus nuclear recoils~\cite{Deap2016}. 
These improvements, as well as a significantly larger active volume, were realised in 2016 with DEAP-3600~\cite{Deap2016} which utilised 3600 kg of liquid argon (of which 1000 kg was within the fiducial region). 
The first completed run of DEAP-3600 ran for 4.44 live days corresponding to 14.8 \tond exposure~\cite{Deap3600} and was able to operate with an even lower energy threshold of 80 photoelectrons (10\keVee) instead of the originally projected 120 PE~\cite{Deap2016}. 
Furthermore, for this run in an energy window of 15 -- 31 \keVee (52 -- 105 \keVnr) no contaminating events were reported after the nuclear recoil event cuts, suggesting a cross-section for 100 \gevc at 90\% C.L. of $< 1.22 \times 10^{-44}$ \cms.

A longer run~\cite{Deap2019} of 231 live days (758 \tond exposure), using 3279 kg of liquid argon, and a small gaseous phase of argon in the upper 30cm of the acrylic cylindrical vessel was recently undertaken. 
In the WIMP region of interest for 95-200 PE range (15.6 - 32.8 \keVee), after all cuts were applied, no background events were observed in accordance with expectations ($0.62^{+0.31}_{-0.28}$)~\cite{Deap2019}.
For WIMP masses above 30 \gevc DEAP-3600 achieved the most sensitive search using a liquid argon target to date, with a spin-independent cross-section of above $3.9 \times 10^{-45}$ \cms for 100 \gevc excluded~\cite{Deap2019}. 
The future for this team is a joint project with the DarkSide collaboration on DarkSide-20k~\cite{DarkSide20k}.

\subsection{Bubble Chambers}
One of the earliest methods of particle detection, pioneered in 1952 by Donald Glasner~\cite{Glaser1952}, bubble chambers contain a liquid in superheated form that, upon reduction in applied pressure, cavitates along the track of any particles traversing the fluid; with the number of bubbles in proportion to the energy deposition of these incoming particles. 
The bubbles expand adiabatically becoming visible to a monitoring camera, enabling the true 3D trajectory of all colliding particles to be tracked. 
An external magnetic field across the chamber ensures that these trajectories, and hence bubble trails, exhibit a helical motion if charged. The radius of this helical motion then determines the charge-to-mass ratio of the original ionised particle as well as incoming velocity~\cite{Glaser1955}. 
Critically, the bubble chamber permits an effectively background free measurement to be made, with obvious benefits to detecting rare collision events such as dark matter.

Modern bubble chambers operate in a similar principle but instead have the superheated liquid distributed into multiple individual droplets within a supporting matrix rather than a single volume driven by an external piston that modifies the pressure. 
In these experiments, individual droplets can be monitored by acoustic sensors to detect nucleation and the resultant acoustic shockwave, upon on the passage of a particle~\cite{Sarkar2017}. 
These experiments are cycled through increasing temperatures to lower the threshold energy sensitivity of the detector and routinely the experiment has to be repressurised to return all bubbles to liquid phase~\cite{Amole2015}.

\subsubsection{PICASSO}
Based in SNOLAB, the PICASSO (Project In CAnada to Search for Supersymmetric Objects) experiment underwent several years of increasing number of $C_{4}F_{10}$ detector elements. 
Ultimately reaching 32 modules containing superheated droplets with total active mass 3 kg~\cite{PICASSO}. 
The experiment used 9 piezoelectric transducers attached to each container wall as acoustic sensors that permitted improved triangulation of events and removal of the more 'active' regions closer to the walls to create a background-free inner fiducial volume.  

The latest PICASSO run concluded in 2014, achieving a final exposure 213.4 \kgd, with competitive limits on lower mass dark matter candidates due to the low mass target nucleus $^{19}F$~\cite{PICASSO}. 
The operating temperature ranged from 303 K to 323 K, equivalent to an energy threshold of 1 -- 40 \keVnr, with no signal above background observed~\cite{PICASSO}. 
Upper 90\% confidence limits of a spin-dependent cross section of $1.32 \times 10^{-2} {\rm pb}$ at 20 \gevc while for the spin-independent case at 7 \gevc dark matter mass, a limit of $4.9 \times 10^{-5} pb$ was reached~\cite{PICASSO}. 
The technologies and techniques developed in PICASSO were combined with COUPP, below, to create the PICO collaboration.

\subsubsection{COUPP}
Similar to PICASSO, the Chicagoland Observatory for Underground Particle Physics (COUPP) experiment also operated in SNOLAB but used $CF_{3}I$ as the superheated target liquid. 
The experiment consisted of a single 150-mm diameter fused silica bell jar monitored by two CCD arrays providing stereoscopic determination of bubble tracks and 4 piezoelectric transducers externally mounted to improve alpha-background discrimination~\cite{COUPP}. 
With 4 kg of superheated liquid, and a total exposure of 437.4 \kgd, the experiment found all signals could be attributed to known background. 

The successor experiment to COUPP, COUPP-60, was later renamed PICO-60 after the formal collaboration with the PICASSO team.

\subsubsection{PICO-60}
Also located at SNOLAB, PICO-60 is a bubble chamber experiment comprised of 52kg of $\rm C_{3}F_{8}$ held in superheated liquid form. 
Optical sensors capture trajectories to discriminate against multiple neutron collisions, alpha decays are removed by their acoustic signals as triggered by piezoelectric sensors~\cite{Amole17}. 
Most recently~\cite{Amole19}, PICO operated for 1404\kgd exposure at 2.45 \keVee, placing  stringent limits on the WIMP-proton spin-\emph{dependent} cross section  of $2.5 \times 10^{-41} {\,\rm cm^{2}}$ for a 25 \gevc WIMP.

A new `right-side up' orientation of the experiment was adopted in PICO-40L~\cite{PICO40} which removed the need for a buffer liquid against the container wall which was the source for significant levels of the contamination. 
It is claimed~\cite{PICO40} that this experimental upgrade will provide a factor 6 improvement on PICO-60 limits~\cite{Amole17}, paving the way for PICO-500, a ton-scale detector with 500l of $\rm C_{3}F_{8}$ fluid at SNOLAB.

\section{Summary}
\label{sec:Summary}

The assumption that dark matter is comprised of a weakly interacting massive particle makes direct detection experiments highly compelling in the search for that particle. 
As a result of basic astronomy there is a predictable flux, and an annual modulation thereof, experienced by terrestrial dark matter detectors which allows for a simple signal to be sought. 
Provided there is \emph{some} interaction mechanism with nuclei beyond gravity  there is a high chance that sooner or later such a signal will be seen. 
Direct detection experiments have proven their capability to scale to larger detector masses, lower backgrounds, or simply wait longer to further explore the available parameter space. 
Despite the enormous success of this approach, a confirmed discovery is still pending.

Searching for an annual modulation in the event rate has the advantage of being less dependent on background reductions and identifications. 
Past experiments have shown the potential of this method and there is the possibility that one experiment has already measured dark matter. 
Several experiments have so far tried to confirm this measurement without success but new ones are already running to further test the claim.

Despite these successes, the future of direct detection experiments is challenging. 
The solar neutrino background presents an unavoidable limit that will be reached in the next generation of facilities. 
This decade will see the direct direction experiments find dark matter or be blinded by the Sun, even underground. 
New technologies have thus to be further explored to ensure a smooth transition, such as directional dark matter detectors~\cite{Mayet2016} like Cygnus~\cite{Cygnus,Cygno}, or alkali halide crystals as bolometers~\cite{Nadeau2014} such as COSINUS~\cite{cosinus2016}. 
The search for dark matter has been long but the collective efforts of nuclear, particle and astrophysicists around the world in the new field of astroparticle physics will surely provide answers to one of the biggest questions in science \cite{NatureBiggestQuestions}.

\section*{Acknowledgements}
The author FF has received funding from the European Union’s Horizon 2020 research and innovation programme under the Marie Sklodowska-Curie grant agreement No 703650.
The authors gratefully acknowledge the work of Swinburne Astronomy Productions’ James Josephides in creation of the `WIMP Wind infographic'.


\printbibliography

\end{document}